\documentclass[a4paper,11pt,amsmath,amssymb,amsfonts,onecolumn,nofootinbib]{article}
\pdfoutput=1
\usepackage{jheppub}
\usepackage[T1]{fontenc}
\usepackage{graphicx}% Include figure files
\usepackage{float}
\usepackage{subcaption}
\usepackage{color}
\usepackage{dcolumn}% Align table columns on decimal point
\usepackage{bm}% bold math
\usepackage{amsmath}
\usepackage{amssymb}
\usepackage{xcolor}
\usepackage{physics}
\usepackage[normalem]{ulem}
\def\d{d\kern-.8 ex\vrule height 1.3 ex depth-1.24 ex width .7 ex \kern .15 ex}
\def\D{D\kern-1.7 ex\vrule height .87 ex depth-.8 ex width .7 ex \kern .95 ex}
\usepackage{xspace}
\newcommand*{\ads}{AdS${}_{4}$\xspace}

\usepackage{bbold}
\newcommand{\be}{\begin{equation}}
\newcommand{\ee}{\end{equation}}
\newcommand{\bea}{\begin{eqnarray}}
\newcommand{\eea}{\end{eqnarray}}

\newcommand{\sbsch}{{\text{Sch}}}

\newcommand{\uv}{\mathrm{UV}}
\newcommand{\ir}{\mathrm{IR}}
\newcommand{\sch}{Schr\"{o}dinger }
\newcommand{\rn}{Reissner-Nordstr\"{o}m }
\def\D{D\kern-1.7 ex\vrule height .87 ex depth-.8 ex width .7 ex \kern .95 ex}

\begin{document}
\title{Emerging Fermi liquids from regulated Quantum Electron Stars}

\author{Nicolas Chagnet${}^1$, Vladan {\D}uki\'c${}^{2,3}$, Mihailo \v{C}ubrovi\'c${}^2$ and Koenraad Schalm${}^1$\\
${}^1${\it
Institute Lorentz for Theoretical Physics, $\Delta$-ITP, Leiden University}\\
{\it Niels Bohrweg 2, Leiden, the Netherlands.}\\
${}^2$ {\it
Center for the Study of Complex Systems, Institute of Physics Belgrade, University of Belgrade, Serbia.}\\
${}^3${\it
Faculty of Physics, University of Belgrade, Serbia.}\\
E-mail: {\tt chagnet@lorentz.leidenuniv.nl}, {\tt cubrovic@ipb.ac.rs}, {\tt vdjukicns@gmail.com}, {\tt kschalm@lorentz.leidenuniv.nl}
}

\abstract{
We construct a fully quantum zero-temperature electron star in a soft-wall regulated anti-de-Sitter Einstein-Maxwell-Dirac theory that is thermodynamically stable compared to the Reissner-Nordstr\"{o}m black hole. The soft wall only acts on the effective mass of the fermionic degrees of freedom, and allows for a controlled fully backreacted solution. The star is holographically dual to an RG flow where a gapped Fermi liquid starts to emerge from a UV CFT, but decouples again once the effective energy scale becomes lower than the the gap of the fermionic degrees of freedom. The RG flow then returns to a non-trivial strongly coupled relativistic fixed point with a holographic dual. Our regulated quantum electron star is thus the fermionic analogue of the Horowitz-Roberts-Gubser-Rocha AdS-to-AdS domain wall solution for the holographic superconductor.
}

\maketitle

% Section 1
\section{Introduction}
Strongly correlated electrons at finite density remain a deep and interesting puzzle, encountered in various quantum-many body systems, from condensed matter to heavy ion physics to astrophysics. %In particular, how do robust and experimentally universal
%non-Fermi-liquid states such as the strange metal arise.
%This is beyond the reach of perturbative physics: 
Apart from some special cases, Fermi liquids are the only interacting fermionic systems at finite density where we have good control. A breakthrough was provided by the application of AdS/CFT to finite density large $N$-matrix fermionic systems. This allowed  new strongly coupled IR fixed points characterized by an emergent Lifshitz scaling with dynamical critical exponent $z$ to be discovered.\footnote{At finite $N$ these fixed points may be not be true IR fixed points but intermediate scale attractors in the RG flow.}  
Though many of such results were found in bottom-up   holographic models where only bosonic operators are tracked, there is reason to believe that any holographic finite density systems must also have microscopic fermionic degrees of freedom. Indeed a number of these holographically discovered fixed points have now been independently confirmed as Sachdev-Ye-Kitaev-like large $N$ quantum spin-liquid fermionic ground states, where the additional microscopic description allows valuable extra insights into the workings of these novel states of matter.

In holography these new ground states are qualitatively understood to arise as a deconfined phase of an underlying microscopic theory with the confined phase corresponding to a conventional Fermi liquid; see \cite{Huijse:2011hp}.
A dozen years ago this was a hotly debated topic and it was found that the prototypical deconfined state, characterized by the AdS$_2$, $z=\infty$ near horizon dynamics of AdS Reissner-Nordstrom (RN) black holes and an associated multitude $N$ 
of non-Fermi-liquid Fermi surfaces \cite{Vegh:2009,Leiden:2009,Faulkner:2009}
%\cite{,Liu:2011,LiuIqbal:2011} 
%indeed 
%transitions 
in the Thomas-Fermi limit of $N\rightarrow \infty$ 
indeed transitions at low temperatures to a charged Tolman-Oppenheimer-Volkov electron star \cite{Hartnoll:es,Larus,Hartnoll:phtr,Hartnoll:2011,Leiden:2011}. These states are partially confined - partially deconfined in that they still have a finite $z$ Lifshitz horizon; for a review  and the transport responses of these states, see  \cite{thebook,Hartnoll:2016apf}.

\begin{figure}[t!]
\begin{center}
\includegraphics[width=0.8\textwidth]{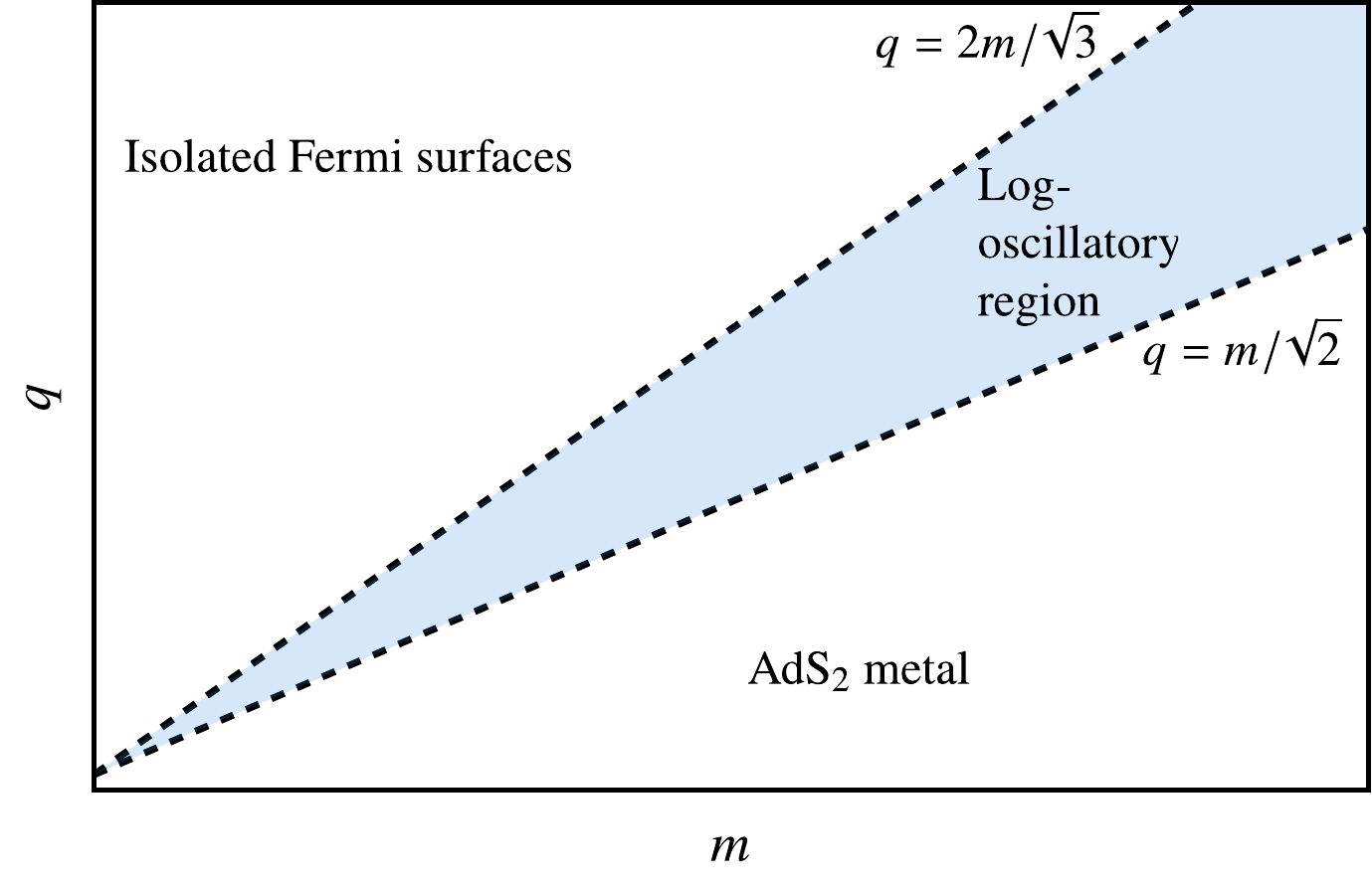}
\caption{A schematic representation of the phase diagram of holographic fermions, where $q$ and $m$ are the charge and the mass (related to the scaling dimension in field theory $\Delta=3/2+m$) of the bulk fermion respectively. Along the line
%$q(m)=\sqrt{2}m$
$q = m/\sqrt{2}$, determined by the Schwinger pair production threshold, the quantum phase transition ought to happen between the \rn black hole describing the strange metal phase and the
quantum electron star solution (no black hole) corresponding to a metallic phase. However, this line is not identical to boundary of the regime where the \rn system supports stable Fermi surfaces as probed through the \rn spectral functions. The electron star (fluid) model requires taking the limit $q,m\to 0$ where both
critical lines become indistinguishable. To understand the transition at finite $q, m$ is the motivation for our approach. Adapted from \cite{Faulkner:2009}.}
\label{figintro}
\end{center}
\end{figure}

However, %despite some impressive progress in the aforementioned directions, 
away from the Thomas-Fermi limit a holographic description of a direct single Fermi-surface
 deconfined non-Fermi-liquid-to-confined Fermi-liquid $T=0$ quantum phase transition has so far not yet been found. %been given a clear answer.  
In the bulk,
this problem corresponds to solving an Einstein-Maxwell-Dirac system in a self-consistent way, accounting for the backreaction of fermions on geometry, but keeping the number of Fermi surfaces finite or specifically keeping only one. %The correspondence between the radial scale in the bulk and the energy scale in
%field theory suggests that the central problem is how the black hole horizon vanishes at the transition point -- this encodes for deep infrared (IR) physics, i.e., low-energy excitations in the CFT, whereas the fixed AdS asymptotics
%lead to the universal high-energy conformal regime in the ultraviolet (UV).
The distinct puzzle here is that the signal of the putative instability towards confinement at low temperature --- a log-oscillatory response in the single fermion spectral function \cite{Faulkner:2009} --- occurs at a distinct point in parameter space from the one where the first stable Fermi surface is located (\autoref{figintro}).
In \cite{Leiden:2013}
an electron star model is introduced where $N$ is finite but still very large; this hinted at a first order rather than a continuous transition. Approaching the question from the other side, a holographic description of confined single Fermi surface Fermi-liquid was constructed in \cite{Sachdev} by enforcing confinement through a hard wall IR cut-off \cite{Sachdev}. This confirmed that confinement-deconfinement is the correct viewpoint of the quantum phase transition, but did not yet include the gravitational backreaction.
%This research is to some extent a continuation of previous work by us \cite{Leiden:2010,Leiden:2011,Leiden:2013} and other authors \cite{Hartnoll:es,Larus,Hartnoll:phtr,Hartnoll:2011,LiuIqbal:2011,Sachdev,McGreevy:2012,McGreevy:2013} in an
% unexpected direction. Those papers aim at a self-consistent solution of the holographic fermion system (within some approximation) and essentially probe different parts of the phase diagram. 
The most comprehensive study to date is the attempt at
quantum electron star model of \cite{McGreevy:2012,McGreevy:2013} %which offers a general
which regulates the system by putting it on a sphere and then tries to carefully remove this regularization procedure for a self-consistent solution of the Einstein-Maxwell-Dirac equations in the asymptotic AdS background. 
% The zero-temperature ground state of
% the system in the large density limit (fluid limit) is constructed and analyzed in \cite{Hartnoll:es,Hartnoll:2011}, under the name of electron star; in this model, the number of occupied fermionic modes is $N=\infty$. In \cite{Leiden:2013}
% an electron star model is introduced where $N$ is finite but very large. These approaches are suitable for studying the high-density limit of the system, deep into the Fermi-liquid-like phase. The idea of \cite{Sachdev} to introduce the IR
% regulator has paved the way for the pioneering studies of the deep quantum regime of the electron star in \cite{McGreevy:2012,McGreevy:2013}, however physical properties of the system like the
% spectrum and the phase transition itself were not studied so far.

The simple hard-wall solution of \cite{Sachdev} already illustrates the fundamental problem. In the presence of an occupied Fermi surface the gravitational backreaction is uncontrolled, see \cite{McGreevy:2012,McGreevy:2013}. These subsequent papers then address this by a second cut-off for the backreaction, and then attempt to remove both cut-offs in a precarious balancing act.
In the present paper we address this in a different way. We construct a fully gravitionally backreacted single-Fermi surface solution confined through a soft rather than a hard wall. From the gravitational point of view this soft wall determines the deep interior boundary conditions of the fermionic wave functions instead of the horizon geometry. As illustrated in detail in \cite{McGreevy:2012,McGreevy:2013} at the technical level the puzzle is that with the vanishing of the horizon (signalling deconfinement) at the quantum phase transition, not only must one find a new self-consistent (confining) IR geometry, but also an associated set of self-consistent boundary conditions for the fermion wave-function.

Because the confining boundary conditions suppress the fermion wave function in the IR, there is also no associated backreaction in the deep IR, which remains AdS. This confined regulated quantum electron star (rQES) is therefore the fermionic analogue of the Horowitz-Roberts-Gubser-Rocha AdS$_4$-to-AdS$_4$ groundstate/domain wall for holographic superconductors \cite{HorowitzRob,Gubser:2008wz}. This solution (just like our soft wall confining electron star solution) describes a system that flows from a conformal pure AdS UV to an intermediate ordered holographic superconductor (Fermi liquid) state with a gap in the sense that below that gap it returns to the renormalized conformal theory and low energy excitations cannot disturb the ordered state. As is well-known the generic holographic superconductor ground state is not AdS$_4$-to-AdS$_4$ but of the Lifshitz type \cite{GubserNellore}. It is the technical difficulties described above that guided us to first construct this Horowitz-Roberts-Gubser-Rocha type solution. We leave the full Lifshitz quantum electron star for future work. One natural way to construct the latter is that, rather than trying to remove the soft-wall regulator, one can also make it dynamical, similar to the electron star study in \cite{HartnollHuijse}.

We do confirm that within the class of non-dynamical soft-wall solutions this gapped confined holographic Fermi liquid is the thermodynamically preferred state over the deconfined \rn metallic state for appropriate charge and mass of the fermion. Because we are not yet able to remove the regulator we do not yet solve the puzzle of \autoref{figintro} directly.

The outline of the paper is the following. In Section \ref{sec:setup} we present the gravity setup and the regulated quantum electron star (rQES) solution. In Section \ref{sec:dualqft}, we present the properties of our rQES solution, i.e. the gapped confined Fermi liquid: we show it is the thermodynamically preferred solution in a
certain range of parameters, and demonstrate the existence of the infinitely long-lived quasiparticle peaks in the spectrum of the boundary theory. In Section \ref{sec:removingReg}, we present some considerations about removing the confining soft wall. Section \ref{sec:conc} sums up the conclusions together with some musings on further directions of work and the physical meaning of our results.

% Section 2
\section{A confined Quantum Electron Star: set-up}
\label{sec:qes}
\label{sec:setup}
The minimal bottom-up gravity dual of a strongly correlated electron system is 
%the minimal bottom-up model which has been studied extensively for several years in the probe limit \cite{Vegh:2009,Leiden:2009,Faulkner:2009} and with some backreaction
%\cite{Leiden:2010,Sachdev}. It is given by 
the Einstein-Maxwell-Dirac system \cite{Vegh:2009,Leiden:2009,Faulkner:2009}.
%, where charged bulk fermions backreact on geometry and the gauge field. 
The new element of our setup is the phenomenological soft-wall-like
regulator inspired by bottom-up AdS/QCD \cite{Herzog:2006ra}. The regulator is a fixed non-dynamic scalar field, which neither backreacts on the metric itself nor does it feel the backreaction by the fermions. This is again in line
with AdS/QCD models. Therefore, the geometry starts as pure AdS in the UV, in the interior it is influenced by the gauge and matter fields and deviates from AdS, and in far IR all
matter fields are exponentially damped by the confining potential. However, in contrast to most hard/soft-wall models we will let the potential only damp the matter sector and not the  gravitational sector.
The action of the system is:
\begin{equation}
	\label{action}
	S = \int \dd^4 x \sqrt{-g} \left[ \dfrac{L^2}{2\kappa^2} \left( R + 6 \right) - \dfrac{L^2}{4} F_{\mu\nu}F^{\mu\nu} + L^3\mathcal{L}_f[\Psi, \Phi] \right]
\end{equation}
where $\kappa$ is the gravitational coupling constant;
%and $F_{\mu\nu} = \partial_\mu A_\nu-\partial_\nu A_\mu$ is the field strength; radial symmetry and the absence of magnetic field imply the gauge field is of the form $A = A_t(z)\dd t$.  Through the rescaling $g_{\mu\nu} \mapsto g_{\mu\nu}L^2, A_\mu \mapsto LA_\mu$, 
%and
and $L$ is set to $L=1$ in the remainder.
The Dirac Lagrangian is:
\begin{equation}
\label{lagfermi}\mathcal{L}_f=\bar{\Psi}\left[e^\mu_A\Gamma^A\left(\partial_\mu +\frac{1}{4}\omega_\mu^{BC}\Gamma_{BC}-i qA_\mu\right)-\left(m + \hat M  \Phi \right) \right]\Psi
\end{equation}
where $\bar{\Psi}=i\Psi^\dagger\Gamma^0$, $e^\mu_A$ is the vierbein, $\Gamma^A$ are the gamma matrices in four dimensions, and $\omega_\mu^{AB}$ is the spin connection.
The regulator is fully encoded in an effective mass contribution $\hat{M}(z)\Phi(z)$ for the Dirac field, with $\Phi(z)$ a non-dynamical scalar field whose profile we shall choose later.
Inspired by \cite{deTeramond:2011qp}, we will consider two types of the confining potential:
\begin{equation}
	\label{matrixmass}
	\hat M = 
	\begin{cases} 
		-e^z_3 \Gamma^3~, &~~ \textrm{the potential preserves  chirality}~,\\
		z \mathbb{1}_4~, &~~ \textrm{the potential breaks chirality}~.\\
	\end{cases}
\end{equation}
Here  $z$, both as index and a variable,  refers to the radial coordinate of the AdS space. We will  assume a radially symmetric metric which is asymptotically AdS${}_{d+1}$ with $d = 3$, parametrized as:
\begin{equation}
\label{metric}
	\dd s^2= - \dfrac{f(z)h(z)}{z^2} \dd t^2 + \dfrac{\dd x_i \dd x^i}{z^2} + \dfrac{\dd z^2}{z^2 f(z)}~.
\end{equation}
The radial coordinate is defined for $z\geq 0$, where $z=0$ is the location of AdS boundary (UV).  Development of a horizon at finite $z$ is in principle signified by the appearance of a zero of the function $f$: $f(z_H)=0$. At zero temperature (the only case we consider), the space extends to infinity, $0\leq z\leq\infty$.

Our choice to let the wall only confine the fermion-matter sector (together with the absence of backreaction by the confining scalar) implies that at finite chemical potential but zero bulk fermion density, the thermodynamically preferred solution is the regular %pure AdS or the 
charged (RN)
 black hole, though pure AdS with a constant electrostatic potential is also a solution. 
%depending on the boundary conditions for the gauge field: requiring zero field strength in the interior leads to pure AdS geometry, whereas zero electrostatic potential $A_t$ requires a nonzero field strength in IR,
% creating the charged RN horizon. In line with the idea to observe the flow away from a quantum critical point, we start from the pure AdS, dual to the UV CFT, and impose the zero field strength as the boundary condition in the IR.

For a certain value of the charge $q$ of the fermion, it will be thermodynamically preferred to store all charge in an occupied bulk fermionic state, i.e. nonzero bulk density $n_c \equiv \langle\Psi^\dagger\Psi\rangle$, rather than a \rn black hole. Now the precise radial profile of the scalar $\Phi(z)$ becomes important. 
% The radial profile $\Phi(z)$ is equally important. Just like for the choice of symmetry, we merely state it here and explain it in full later on. 
The original AdS/QCD papers used a quadratic scalar, behaving in the IR as $\Phi\sim z^2$ \cite{Karch:2006pv}, which ensures confinement while still being smooth. Another form found in the literature is a profile which flattens out to a constant in the IR \cite{Fang:2016uer}. At the same time the UV completion of the scalar field has to ensure that its contribution to the Dirac equation decays quickly enough for small $z$ to reproduce the equation of motion in pure AdS in the limit $z \to 0$. The forms that satisfy all the requirements and which we find numerically convenient are
\begin{equation}
	\label{eq:scalarProfiles}
	\begin{aligned}
		\Phi(z)&= \lambda z^2,&~~ &  \textrm{quadratic scalar}\\
		\Phi(z)&= \lambda\frac{z^\alpha}{z_0^\alpha + z^\alpha},& ~~ & \textrm{flat scalar}.
	\end{aligned}
\end{equation}
The amplitude of the scalar (i.e. the measure of the "hardness" of the wall) is parametrized by $\lambda$, and $z_0$ is the scale at which the scalar begins to flatten (in the second, flat scalar model). The choice of $\alpha$ is merely that of computational convenience and we choose $\alpha = 4$. Similarly, we will consistently choose $z_0 = 2$ throughout the rest of this paper.

\subsection{Einstein-Maxwell-Dirac equations}
From the action we obtain the Maxwell equation and two convenient linear combinations of the $tt$ and $zz$ components of the Einstein equations. With the ansatz that only $A_t\neq 0$, and that all functions only depend on $z$, compatible with homogeneity and isotropy, they reduce to
\begin{equation}
	\label{eomBackgrounds}
	\begin{aligned}
		A_t''(z)-\dfrac{h'(z)}{2h(z)}A_t'(z) ~& = ~\sqrt{h(z)} n(z)~,\\
		1+\dfrac{z}{3}f'(z)-f(z) ~& =~ \dfrac{z^2}{3f(z)h(z)}\rho(z)+\dfrac{z^4}{12 h(z)}A_t'(z)^2~,\\
		h'(z) ~& =~ -zh(z)p(z)-\dfrac{z}{f(z)^2}\rho(z)~.
	\end{aligned}
\end{equation}
Compatible with the symmetries the current vanishes $J^i=0$, the charge density $J^0$ is denoted as $J^0=n(z)/\sqrt{-g} = z^4 n(z)/\sqrt{h(z)}$, and the stress tensor is parametrized as 
\be
\label{tmunu}(T_f)_{\mu\nu}=\mathrm{diag}(\rho(z),p_\perp(z),p_\perp(z),p(z)),
\ee
where $p_\perp(z)$ is the pressure in the transverse $x,y$ directions. %The $ii$ component ($i=x,y$) of the Einstein equations is not independent of the previous two and can be dropped; hence $p_\perp$ does not appear [HOW DOES THIS WORK]?}.\\
%\nc{Careful, the proper definition of the stress tensor for these equations is
%\be
%(T_f)_{\mu\nu}=\mathrm{diag}(\rho(z),p_\perp(z),p_\perp(z),p(z)),
%\ee
%[KS PLEASE CHECK THIS CAREFULLY]
%}\\

The $ii$ components of the Einstein equations are both equal to
\begin{align*}
    z h(z) \left[-z^3 A_t^\prime(z)^2+\left(3 z f^\prime(z)-4 f(z)\right) h^\prime(z)+2 z f(z) h^{\prime\prime}(z)\right]+ &\\
    +2 h(z)^2 \left[z \left(z f^{\prime\prime}(z)-4 f^\prime(z)-2 \beta  z p_\perp(z)\right)+6 f(z)-6\right]-z^2 f(z) h^\prime(z)^2  & = 0~.
\end{align*}
They are not independent, however. Denoting the Einstein field equations as $E_{\mu\nu} \equiv G_{\mu\nu} - T_{\mu\nu}$ and the Maxwell equation as $E_M \equiv \nabla_\mu F^{\mu\nu} - J^\nu$, one can show that\footnote{This is essentially $\nabla_{\mu}G^{\mu\nu}=\nabla_{\mu}T^{\mu\nu}$.}
% observe that, provided the ansatz \eqref{metric}, $E_{tt}$ and $E_{zz}$ are two first-order ordinary differential equations in $\{f, h\}$. The two other terms are $E_{xx} = E_{yy}$ by rotation invariance, which form a second order equation in both $f$ and $h$. It is quite easy to show that we have
\be
    E_{xx} = \hat L \cdot E - \dfrac{1}{2 z} \nabla_\mu T^{\mu\nu}~,
\ee
where $\hat L \cdot E \equiv A_1 \partial_z E_{tt} + A_2 \partial_z E_{zz} + A_3 E_M + A_4 f^\prime(z) E_{tt} + E_{zz} \left( A_5 f^\prime(z) + A_6 h^\prime(z) + A_7 \right)$ is a linear combination of both $\{E_{tt}, E_{zz}, E_M\}$ and their derivatives and $T^{\mu\nu}$ is the total stress-energy tensor associated with the matter content of the theory. The stress-tensor is covariantly conserved if the matter sector is on-shell, i.e. obeys its equations of motion. Thus
\be
    E^{\mathrm{on-shell}}_{xx} = \nabla_\mu T^{\mu \nu} = 0 
\ee    
It is therefore sufficient to solve the three equations \eqref{eomBackgrounds} together with the matter sector.

The charge, energy and pressure densities $n(z),\rho(z),p(z)$ are %respectively the fermionic density, energy density and longitudinal pressure. 
determined by the occupied fermionic states in the AdS bulk space.
Importantly, we will compute them %(\ref{tmunu}) is
\emph{solely %determined by the
from microscopic considerations}: we do \emph{not} assume anything like a fluid limit or a specific form of
the equation of state. We compute them from the Dirac Lagrangian, within the one-loop Hartree correction to the background. This is discussed in detail in the next subsection.

We will now proceed to derive the equation of motion for the Dirac field. From (\ref{lagfermi}), the equation reads:
\begin{equation}
\label{diraceq}e^\mu_A\Gamma^A\left(\partial_\mu+\frac{1}{4}\omega_\mu^{BC}\Gamma_{BC}-iqA_\mu\right)\Psi=\left( m + \hat M(z) \Phi \right) \Psi~.
\end{equation}
It is known that the spin connection in this type of metric can be eliminated by rescaling the fermion \cite{Vegh:2009,Iizuka:2011hg}:
\begin{equation}
\label{spincon}\Psi=\left(-g^{zz}\det g_{\mu\nu}\right)^{-\frac{1}{4}}\tilde{\psi}=\left(\frac{f(z)h(z)}{z^{2d}}\right)^{-\frac{1}{4}}\tilde{\psi}\equiv a(z)\tilde{\psi}.
\end{equation}
In addition, it is convenient to eliminate any singular terms from the fermionic wavefunction. Since our solutions are smooth in the interior as we shall see, the only singularity is the branch cut in the UV behaving as $z^m$. We
thus rescale one more time
\be
\label{rescaling}\tilde{\psi}=z^m\psi\equiv b(z)\psi.
\ee
In most cases we will use the rescaled form and write the equations for $\psi$. So far this is all independent of the gamma matrix representation. In order to simplify the equations of motion, we now employ the representation
\begin{equation}
\label{repka}\Gamma^\mu=\left(\begin{matrix}0 & \gamma^\mu\\ \gamma^\mu & 0\end{matrix}\right),~~\Gamma^3=\left(\begin{matrix}1 & 0\\ 0 & -1\end{matrix}\right),
\end{equation}
with $\mu \in \{0, 1, 2\}$, $\gamma^0=i\sigma^2,\gamma^1=\sigma^1,\gamma^2=\sigma^3$ and $\sigma^{1,2,3}$ are the usual Pauli matrices. Homogeneity and isotropy along the $t,x,y$ directions allow us to take the energy $\omega$ and momentum $k\equiv k_x$
as good quantum numbers, so the Dirac bispinor is expressed as
\be
\psi=e^{-i\omega t+ i kx}\left(\psi_1(z),\chi_1(z),-i\chi_2(z),i \psi_2(z)\right)^T.
\ee
As in \cite{Sachdev,Iizuka:2011hg}, this yields two (equivalent) decoupled systems for the two independent components, for $\psi_{1,2}$ and $\chi_{1,2}$, corresponding to the spin degeneracy of our system. We will focus on the $\psi_i$ components for which the Dirac equation reads
\begin{equation}
	\label{diraceqfin}
	\begin{aligned}
        \left[\partial_z + \varepsilon_+ \Phi+\dfrac{m}{z}\left(1-\frac{1}{\sqrt{f(z)}}\right)\right]\psi_1(z)-\left[\dfrac{k}{\sqrt{f(z)}}+\dfrac{\omega+qA_t}{f(z)\sqrt{h(z)}}\right]\psi_2(z) & = 0\\
    	\left[\partial_z + \varepsilon_- \Phi +\dfrac{m}{z}\left(1+\frac{1}{\sqrt{f(z)}}\right)\right]\psi_2(z)+\left[\dfrac{\omega+qA_t}{f(z)\sqrt{h(z)}}-\dfrac{k}{\sqrt{f(z)}}\right] \psi_1(z) & = 0\, .
	\end{aligned}
\end{equation}
where $\varepsilon_+ = \varepsilon_- = 1$ corresponds to the chiral-preserving potential and $\varepsilon_+ = - \varepsilon_- = -1/\sqrt{f(z)}$ corresponds to the chiral-breaking potential.
% Section 2.2
\subsection{Fermion densities and backreaction}
The fermionic densities and pressures are obtained microscopically, from the Dirac Lagrangian (\ref{lagfermi}):
% Now we can evaluate the stress-energy tensor. It is obviously diagonal for symmetry reasons; likewise, due to the isotropy in the $x-y$ plane, we will have $T_{xx}=T_{yy}$. We thus expect the stress tensor of the form (\ref{tmunu}), with the energy density and the pressures as the only components. For the energy density $\rho$ and the charge density $n$ we find by definition, differentiating the 
% Lagrangian with respect to $g_{tt}$ and $A_t$:
\begin{eqnarray}
\label{rho1}\rho&=&\langle\Psi^\dagger e_0^t\Gamma^0(-i\omega-i q A_t)\Psi\rangle~, \nonumber \\
\label{n1}n&=&-\langle\Psi^\dagger\Psi\rangle~.
\end{eqnarray}
The components of the pressure $p_\perp,p$ are likewise formally equal to
\begin{eqnarray}
\label{p1}p_\perp&=\langle\bar{\Psi}i e_1^x k_x \Gamma^1\Psi\rangle,\nonumber\\
p&=\langle\bar{\Psi}e_3^z\Gamma^3\partial_z\Psi\rangle.
\end{eqnarray}
The expectation value $\langle\ldots\rangle$ in (\ref{n1}-\ref{p1}) is  the quantum-mechanical expectation value, i.e. one solves the Dirac equation with appropriate boundary conditions (see below) and sums over the quantum numbers in the appropriate range. The quantum numbers are the radial modes $\ell$, and momenta $k_x, k_y$ in the $x,y$-directions which determine the on-shell energy in terms of a dispersion relation $\omega=E_{\ell}(k)$. The role of the confining potential is essential here: it quantizes the radial
number $\ell$. Each discrete radial mode corresponds to a separate Fermi surface \cite{Vegh:2009,Leiden:2009,Faulkner:2009,Hartnoll:2011,Leiden:2011,Sachdev}. As emphasized in the Introduction, we seek a state where only a single Fermi surface is occupied. This must be the lowest radial mode. Note that despite occupying a single mode, this mode still contains a thermodynamically large number of states counted by the $x,y$-momenta. Each radial mode is thus a fluid of fermions.

We will ignore the subtleties of the zero-point energy and the Dirac sea; in principle these are absorbed in a renormalization of the cosmological constant and the AdS radius; see however \cite{McGreevy:2012,McGreevy:2013} for a more detailed treatment.
Then, in terms of the solutions to the Dirac equation, formally the expressions for the density are
\begin{align}
	\label{eq:thermoQuantities}
	n(z) & = \dfrac{2 q}{z^3 \sqrt{f(z)}} \, a(z)^2b(z)^2\sum_{k,\ell}\Theta \left(-E_{\ell}(k)\right)~\left(\psi_{1;\ell,k}^\dagger(z) \psi_{1;\ell,k}(z)+\psi_{2;\ell,k}^\dagger(z) \psi_{2;\ell,k}(z)\right) \nonumber \\
	\rho(z) & = a(z)^2b(z)^2e^t_0(z)\left(- i \omega - i q A_t(z)\right)\sum_{k,\ell} \Theta \left(-E_{\ell}(k)\right)\left(\psi_{1;\ell,k}^\dagger(z) \psi_{1;\ell,k}(z)+\psi_{2;\ell,k}^\dagger(z) \psi_{2;\ell,k}(z)\right) \nonumber \\
	p(z) & = a(z)^2b(z)^2e^z_3(z)\sum_{k,\ell} \Theta\left(-E_{\ell}(k)\right)\left(\psi^\dagger_{1;\ell,k}\partial_z\psi_{2;\ell,k}-~\psi^\dagger_{2;\ell,k}\partial_z\psi_{1;\ell,k}\right)\, 
\end{align}
where the step-function $\Theta(x)$ selects the positive energy states.
Note that due to the antisymmetry of the two spin components, the derivatives of the scaling factors $a(z), b(z)$ cancel out in the expression for $p$.

\subsubsection{The self-consistent Hartree calculation}

We solve the system (\ref{eomBackgrounds}, \ref{diraceqfin}) in the one-loop Hartree approximation. As a reminder, the Hartree correction is the local single-particle diagram (vacuum bubble), ignoring anti-particles, i.e. ignoring the contribution from the Dirac sea. 
%which does not depend on the statistics (Bose or Fermi). 
We do not take into account the Fock correction. 
%the non-local single-particle exchange diagram. 
In flat space, the Hartree correction is trivial \cite{Landau9}: in terms of the causal fermionic propagator $G_R$ it equals $\lim_{t\to 0-}\int \dd \omega \dd^2 k~G_R(\omega,k)e^{-i\omega t}=\delta\mu$,\footnote{The infinitesimal time separation $t\to 0-$ is really the point-splitting regularization, as the integral of $G_R$ at coincident points in spacetime generally diverges; the sign of $t$ is dictated by the contour choice for the retarded propagator \cite{Landau9}.} merely renormalizing the chemical potential. In curved space however, the local chemical potential is $\mu_\mathrm{loc}(z)=A_t(z)\sqrt{-g^{tt}(z)}$, with a nontrivial radial profile, thus the correction $\delta\mu(z)$ is also variable along $z$ and therefore it can have nontrivial physical effects.

\begin{figure}[t!]
	\center
	\includegraphics[scale=0.4,angle=90]{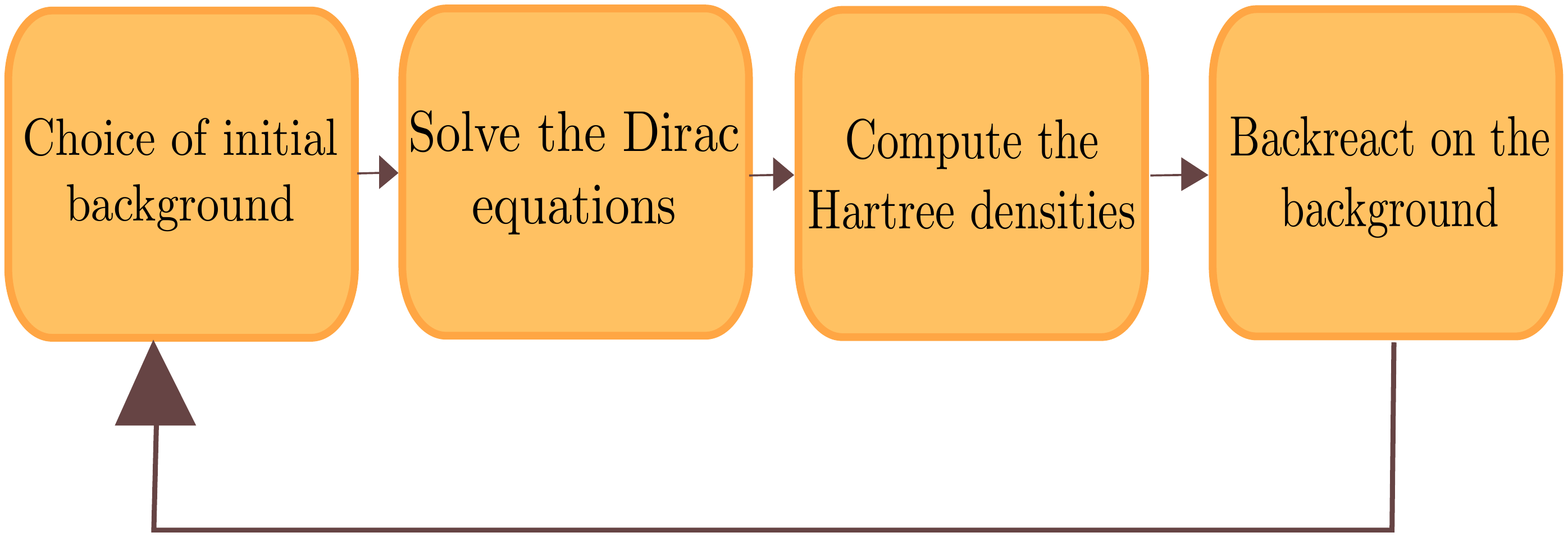}
	\caption{Iteration algorithm used to compute the rQES solution.}
	\label{fig:algo}
\end{figure}

The Hartree approximation then proceeds by computing this one-loop Hartree correction self-consistently. One starts with an ansatz for the background, solves the Dirac equation in this background, computes the one-loop Hartree densities in the assumption that they are small, updates the background and iterates to convergence as in Fig.\ref{fig:algo}.

% The Hartree summation is performed over all radial modes $\ell=1,2,\ldots$ with non-empty Fermi seas, so we find the solution as a discrete sum over the levels: $\psi=\sum_{\ell=1}^{\ell_\mathrm{max}}\psi_\ell$. In addition to $\ell$, we also have the continuous quantum numbers $\omega,k$ for every $\ell$. The normalizable modes (which correspond to field theory states according to the AdS/CFT dictionary) for which the boundary states are that of a Fermi gas define the dispersion relation $\omega=E_\ell(k)$, where $E_\ell(k)$ is the energy eigenvalue.\footnote{Since the bulk Fermi sea consists solely of bound states, with negative energies, it is convenient to consider the negative energy value. This is also in line with the notation of \cite{Sachdev}.} This is equivalent to requiring that the Feynman Green's function have a pole along these modes as we show from the boundary conditions.
% The iteration algorithm used to build our rQES is thus the following:
% \begin{enumerate}
% 	\item Fix an ansatz for the background fields $f,h,A_t$ (in the first iteration pure AdS: $f=h=1,A_t=\mu=\mathrm{const.}$).
% 	\item Solve the Dirac equations \eqref{diraceqfin} to get $\psi_{1,2}(\ell;E_\ell(k),k;z)$.
% 	\item Compute the currents \eqref{thermoQuantities} from the wavefunctions $\psi_{1,2}$.
% 	\item Backreact on the background fields through equations \eqref{eomBackgrounds}.
% 	\item Go back to step 2 and repeat until convergence is reached.
% \end{enumerate}
% The iteration algorithm used to build our rQES is shown in \autoref{fig:algo}.

% Section 2.3
\subsection{Boundary conditions on the Einstein-Maxwell sector}
The Einstein-Maxwell equations (\ref{eomBackgrounds}) require four boundary conditions in total (two for $A_t(z)$ and one for each of the metric functions $f(z), h(z)$). %As usual, physical requirements impose more than four conditions, so some
%of them can only be implemented indirectly, by landing the solution onto the correct branch. 
The UV boundary conditions are
%\footnote{In principle, one could work in the canonical ensemble instead of grand canonical, which would mean fixing the total charge $Q=-A_t'(z_{\uv})$. This choice however poses some technical problems when integrating the equations of
%motion, so we leave it for further study and work exclusively in the grand canonical ensemble.}
\begin{equation}
	\label{bndUV}
	\begin{aligned}
		A_t(z_{\uv}) & = \mu~, & ~~ & \textrm{the chemical potential.}\\
		f(z_{\uv}) & = h(z_{\uv}) = 1~,& ~~ & \textrm{\ads asymptotics}.\\
%		\dfrac{\psi_2(z_{\uv})}{\psi_1(z_{\uv})} & = -z_{\uv}\dfrac{\omega +q\mu-k}{2m+1}~,& ~~ & \textrm{normal modes.}
	\end{aligned}
\end{equation}
The fourth boundary condition we impose is given by our demand that we seek a state where {\em all} the charge is contained in occupied fermionic states.\footnote{There could be interpolating solutions with both a charged horizon and a charge in occupied fermionic states. We will not seek for those here as the presence of the charged Reissner-Nordstrom like horizons should imply the continued presence of log-oscillatory instabilities.}  The confining potential ensures that the fermionic wavefunctions are localized at a finite value in the radial direction. Thus by construction the charge density will vanish in the deep AdS interior. From this follows that the fourth boundary condition is $\partial_z A_t(z_{\text{IR}}) =0$.
%\ks{[TRUE?] \nc{Yes it is the BC we impose.}}. 
Formally $z_{\text{IR}}=\infty$; in our numerical computation it will be finite but large, and we have checked that our results do not depend on its value.
%\ks{[TRUE?] \nc{Yes}}

In practice, we solve the boundary value problem by shooting from the IR. We impose directly the condition $\partial_zA_t(z_{\ir})=0$ as well as the condition $\partial_zf(z_{\ir})=0$. The latter indirectly encodes our demand that we seek a $T=0$ solution; recall that for a black hole solution $\partial_zf(z_{\text{horizon}}) \sim T$. Then we use the free value $A_t(z_{\ir})$ and $h(z_{\ir})$ to shoot for $A_t(z_{\uv})=\mu,~h(z_{\uv})=1$ at the boundary. From the equation of motion for $f(z)$ one obtains automatically that $f(z_{\ir})=1$ once we fall on the right branch; for the same reason one can also use $f(z_{\ir})=1$ as an IR boundary condition if one demands in addition that there is no energy density or electric field in the deep interior.

% Section 2.4
\subsection{Boundary conditions for the fermions}
The UV boundary conditions for the appropriate solutions to the Dirac equation are straightforward.
Near the AdS boundary the rescaled field behaves as
\begin{equation}
\label{bndpsi}
	\begin{aligned}
		\psi_1(z\to 0) & \sim A_{\ell}(\omega,k) \frac{\omega -k-\mu q}{2m-1} z^{1-2m} + B_{\ell}(\omega,k) +\ldots,\\
		\psi_2(z\to 0) & \sim A_{\ell}(\omega,k) ~ z^{-2 m}+ B_{\ell}(\omega,k) \frac{\omega + k-\mu q}{2m+1} z+\ldots~.
	\end{aligned}
\end{equation}
On-shell solutions are normalizable, i.e. $A_{\ell}(\omega,k)=0$. This agrees with 
the AdS/CFT dictionary, where a finite $A_{\ell}(\omega,k)$ would imply an external source for the fermions for a specific band $\ell$ and energy $\omega,k$. Demanding normalizability $A_{\ell}(\omega,k)=0$ instead, implicitly translates in a dispersion relation $\omega(k) = E_\ell(k)$. 
%In fact, the absence of the imaginary part in the energy (i.e., $E_\ell(k_F)=0$ instead of $E_\ell(k_F)=0+\imath(\ldots)$) means we actually have normal, not quasinormal modes, as could indeed be expected at zero temperature.

The IR boundary conditions for the fermions %tend to be subtle and
require a more detailed discussion. 
%A zero-temperature geometry free of singularities determines the behavior of $f$ and $h$. The behavior of the electrostatic potential is wired into the Gauss-Ostrogradsky
% theorem (flux conservation) \cite{Sachdev}: the total flux of the electric field sourced by the bulk matter (Dirac fermion, in our case) equals the total flux through the IR and UV boundary. Since the latter is just the total CFT
% charge $Q$, we can write it as
% \be
% \label{go} Q = Q_\mathrm{f}+\partial_z A_t(z\to\infty).
% \ee
% In other words, if the charge is carried solely by the fermions, indicating a coherent system in the language of \cite{Gouteraux:2011ce}, the electric field strength vanishes in deep IR. As we have discussed in the introduction, we are now 
% interested precisely in such a system, even though, in the long term, the regime where the charge is shared between the bulk and the horizon is the most physically relevant and puzzling. 
Firstly, for the fermionic wavefunctions, the amplitude is set by normalization of each wavefunction to unity. 
% The normal modes exist for any radial quantum number $\ell$ but only on discrete shells in (transverse) momentum space, corresponding to the position of the quasiparticle pole in the propagator. A normal mode defines a bound state in the bulk (finite-density state in the dual field theory), thus it must have a finite norm, meaning that it should fall off sufficiently quickly in the interior. Therefore the eigenenergies are determined by the condition
For each radial mode $\ell$ this implies
\begin{equation}
\label{bndpsieigen}\int dz\sqrt{-g}\vert\psi_{i;\ell,k}(z)\vert^2<\infty.
\end{equation}
For finite temperature backgrounds this is usually not an issue as the horizon is parametrically at finite distance and finite IR boundary conditions, together with the UV-condition that the un-normalizable fall-off vanish, guarantees a finite integral. For the $T=0$ background we consider here, the interior is parametrically at infinite distance and finiteness of the integral can only follow from bounded behavior of the wavefunction.
Since the spin components are not independent, it is sufficient to demand $\psi_{1;\ell}(z\to\infty)\to 0$, i.e., the leading component should vanish in the interior. 
% Note that in the presence of a horizon the integral is over a finite range and the normalization condition is less stringent. We purposely seek the $T=0$ solution without a horizon, hence the normalization condition is important. 

It is well known in AdS/CFT that it is then the simultaneous requirement of a UV and an IR boundary condition that determines the spectrum of the small excitations.  This
spectrum can still be continuous or discrete; we address this directly below. Formally, however, the normalization together with two boundary conditions make the system overconstrained and one must search for accidental solutions. We again do so by shooting from the interior to search for parameters where the UV conditions are also satisfied. 

The shooting condition we use is the ratio $\psi_2/\psi_1$, which still leaves the freedom to normalize the norm (\ref{bndpsieigen}) to unity, and which we do after the solution is found.

% \mc{For all nonzero k the bnd conds dont change. But for k=0 zero mode the bnd conds might be special. And this mode is the only thing that remains when right at crit point and kf=0.}

\subsubsection{Effective potentials and confinement}

Pure $T=0$ AdS --- representing a deconfined phase of the strongly coupled boundary theory  --- has a
continuum spectrum of normal modes computed in the way described above. The system must be considered in a different phase or have its IR dynamics modified by a confining potential to discretize the spectrum; this spectrum may still be ungapped or gapped.
We will now demonstrate that the chiral-breaking soft-confining potential supports a discrete Fermi surface, i.e. a tower of bound states at discrete energies, for momenta up to some $k_F$, the Fermi momentum. The spectrum is also gapped. A convenient way to see the
effect of this potential is to transform the Dirac equation to the \sch form
%, as it is traditionally done in AdS/CFT
\cite{Faulkner:2009,Leiden:2011,thebook}:
\bea
	\label{schro}
	%&\chi_\sbsch(z) = e^{\frac{1}{2} \int_0^z \dd u \, \mathcal{Q}(u)/\mathcal{P}(u)} \psi(z)~,  \nonumber \\
	&\chi_\sbsch(z) = e^{\frac{1}{2} \int_0^z \dd u \, \mathcal{P}(u)}~,  \nonumber \\
	&\left[\partial_z^2 - V(z)\right] \chi_\sbsch(z) = 0\, , \nonumber \\
    %&V(z) = \dfrac{\mathcal Q(z)^2 - 2 \mathcal Q(z) \mathcal P^\prime(z) + 2 \mathcal P(z) (-2 \mathcal R(z) +\mathcal Q^\prime(z) )}{4 \mathcal P(z)^2}~,
    &V(z) =\dfrac{1}{2}\mathcal P^\prime(z)+ \dfrac{1}{4} \mathcal P(z)^2  -\mathcal Q(z) ~,
\eea
where the coupled equations \eqref{diraceqfin} were decoupled into two second order equations, each taking the form
\be
    \psi^{\prime\prime}(z) + \mathcal P(z) \psi^\prime(z) + \mathcal Q(z) \psi(z) = 0~,
\ee
with the indices 1,2 on $\psi, \chi_\sbsch(z), V$ omitted.

\begin{figure}[t!]
    \centering
    \includegraphics[width=\textwidth]{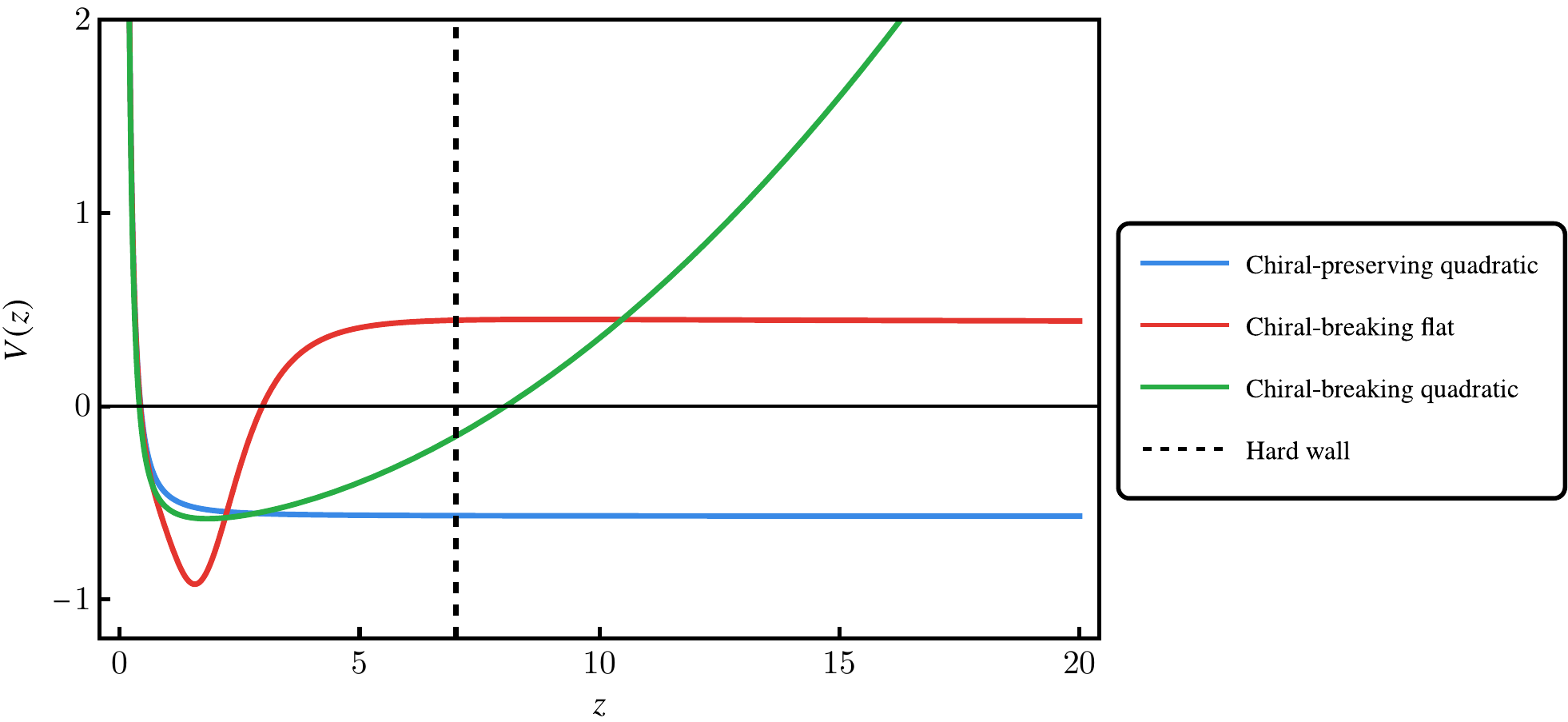}
    \caption{Comparison of the \sch potentials for $\psi_1(z)$ for the two types of confining potential: chiral-breaking quadratic (green), chiral-breaking flat (red) and chiral-preserving quadratic (blue). The dashed black line indicates the truncation of spacetime which
    happens in the hard wall model of \cite{Sachdev} at $z = 7$. Only the chiral-breaking potential and the hard wall allow for bound states. Parameters are $\{m, \mu q, k, \omega\} = \{0.1, 1.05, 0, -0.027\}$. The scalar parameters are $\lambda = 0.1$ for the two quadratic scalars and $\lambda = 1$ for the flat scalar.}
    \label{fig:potentials}
\end{figure}

In principle, the \sch
potential is itself a function of the background spacetime and electrostatic potential $f(z), h(z), A_t(z)$ and can be fully determined only by calculating numerically the full solution. However, we can give a qualitative estimate whether it is confining or not by studying its asymptotics. Since the bulk remains asymptotically \ads, we have $V(z \to 0) \sim \dfrac{1}{z^2}$. In pure \ads the IR behavior would be $V_{\text{AdS-IR}}(z \to \infty) = - (\omega + \mu q)^2 + k^2 + m(m+1)/z^2 + \mathcal O(1/z^3)$ (\autoref{fig:potentials}).\footnote{We are interested in $k^2 < (\omega + \mu q)^2$ since the potential is otherwise confining even in \ads with no regulator, as discussed in \cite{Gubser:2009dt}. We will discuss this later.}
%[CHECK \nc{done}]
This now gets modified by the confining potential due to the scalar $\Phi(z)$. Making the ansatz that the confining potential in the deep IR for $z\to \infty$ suppresses exponentially all sources in the Einstein and Maxwell
 equations for large $z$, i.e. the geometry in the deep IR is again an (emergent) \ads geometry, the leading order IR behavior of the potential is then schematically 
%[CHECK \nc{done}]

% Crucially, the contribution of the \emph{chiral-preserving regulator} to the \sch potential is vanishing. To see this, we can compute the \sch potential while leaving the relative sign $\epsilon \in \{-1,1\}$ between the two
% contributions arbitrary, i.e., the regulator enters the equations as $\Phi(z) \psi_1(z)$ and $\epsilon \Phi(z) \psi_2(z)$ . The leading order of the potential in the infrared is then schematically

\begin{equation}
	 V_{\mathrm{AdS-IR}}=V(z\to\infty) +  (\varepsilon_- - \varepsilon_+) \left[- \frac{\phi^\prime(z)}{2z} + \dfrac{\phi(z) (4m + 2)  +(\varepsilon_- - \varepsilon_+)\phi(z)^2}{4z^2}   + \right] + \mathcal O(1/z^3)~,
\end{equation}

%\ks{[CHECK WHICH $\varepsilon$ FROM Eq. (2.16) \nc{done}]}
%where $\dots$ corresponds to terms not involving $\Phi$. 
Note that the chiral-preserving solution $\varepsilon_+ = \varepsilon_- = 1$ leads to a vanishing contribution and therefore does not lead to  fermionic bound states. In contrast the chiral-breaking solution $\varepsilon_+ = - \varepsilon_- = -1/\sqrt{f(z)} = -1 + {\cal O}(1/z)$ in an \ads IR does lead to a potentially bounding potential depending on the choice of $\Phi(z)$. For this reason, we will work solely with the chiral-breaking scalar field. %Sketching the result in Fig.\ref{fig:potentials}, this also explains our choices Eq.\eqref{eq:scalarProfiles} for the profile $\Phi(z)$.

%Now we can directly study the existence of the potential well for the various regulator choices. Since we are looking for a discrete spectrum for the fermions, we require the \sch potential to be confining. 

\autoref{fig:potentials} shows the behavior of the \sch potential for the various profiles of the scalar field and regulation schemes. %The chiral-preserving regulator at leading order yields a non-confining flat potential in the IR,
%just like in pure AdS with no regulator, in accordance with the above considerations. 
With a chiral-breaking regulator, we indeed see that the infrared behavior of the potential is dominated by the large $z$ behavior of each
profile. The final choice of which scalar field profile to use is determined by the convergence of the iteration scheme. We numerically found the quadratic profile to be unstable while the flat profile leads to an emergent \ads in the
infrared. Specifically for the {chiral-breaking confining potential with flat asymptotics} the \sch potential in the deep IR becomes
\begin{equation}
	\label{eq:IRSchroAdS4}
	V(z \to z_{\ir}) {=} -\omega^2_{\ir} + \lambda_{\ir}^2 + k_{\ir}^2 + \mathcal{O}(1/z) \equiv V_{\ir} + \mathcal{O}(1/z) \, ,
\end{equation}
%\ks{[THERE IS THE CONSTANT $-c$ MISSING HERE, NO? \nc{There is no constant, the leading order is given by omega, k} ]}
where we have used that $f(z), h(z), A_t(z)$ become constant in the emergent \ads IR %values 
and we have defined $\omega_{\ir} \equiv \dfrac{\omega + q A_t(z_{\ir})}{f(z_{\ir}) \sqrt{h(z_{\ir})}}$, $\lambda_{\ir} \equiv \dfrac{\lambda}{\sqrt{f(z_{\ir})}}$ and $k_{\ir} \equiv \dfrac{k}{\sqrt{f(z_{\ir})}}$.\\

In the IR limit, the \sch equation becomes
\begin{equation}
\label{IRexp}
	\left[ \partial_z^2 - V_{\ir} \right] \chi_{\text{Sch}}(z) = 0 ~,
\end{equation}
which is solved by
\begin{equation}
	\label{solFlat}
	\begin{aligned}
		\chi_{\text{Sch}}(z) & = \chi_{\text{Sch}+}(z) e^{\sqrt{V_{\ir}} z} + \chi_{\text{Sch}-}(z) e^{-\sqrt{V_{\ir}} z} \, .
	\end{aligned}
\end{equation}
We see from \eqref{solFlat} that, for frequencies such that $V_{\ir} > 0$, the solutions have a growing and a decaying branch.
% and we will keep the decaying one to select the normal modes.
The decaying branch clearly confines the wavefunction. This is the one we shall choose. This leads to the following IR form for our original Dirac fermion components
\begin{equation}
	\label{eq:IRbcFW}
	%\begin{aligned}
		\psi^{\mathrm{IR}}_{1,2}(z) = c_{1,2}^{\mathrm{IR}}(z) e^{-\sqrt{V_{\ir}} z} \, ,\\
		%\psi^{\mathrm{IR}}_2(z) & \propto e^{-\sqrt{V_{\ir}} z} \, .
	%\end{aligned}
\end{equation}
where the ratio of the coefficients is fixed by the Dirac equation \eqref{diraceqfin}:
\begin{equation}
	\dfrac{\psi^{\mathrm{IR}}_2(z)}{\psi^{\mathrm{IR}}_1(z)}=\frac{c_2^\mathrm{IR}(z)}{c_1^\mathrm{IR}(z)} = \dfrac{1}{\omega_\ir + k_\ir} \left[ \dfrac{m}{z} \left(\dfrac{1}{\sqrt{f_\ir}} - 1\right) + \sqrt{V_\ir} + \lambda_\ir\right]
\end{equation}
and the normalization of the wavefunction to unity sets the remaining overall scale.\\
With these IR boundary conditions the equations \eqref{diraceqfin} are solved by shooting from $z_{\ir}$ to $z_{\uv}$.
% The analytical solutions $\psi_{1,2}^{\mathrm{IR}}$ can then be used as IR boundary conditions to solve the equations (\ref{diraceqfin}) from $z_{\ir}$ to $z_{\uv}$. 

The confinement imposed by both IR and UV boundary conditions leads to a discrete and gapped spectrum which defines a band structure (see \autoref{fig:fitBand} later). The  fall-off of the wavefunction both at the AdS boundary and the interior also implies an absence of any backreaction in those regions. Once backreaction is included the resulting  solutions will therefore be \ads-to-\ads domain wall solutions, as we will show in the next section.
% \textbf{MC The rest of this paragraph to the next section?} \nc{\sout{The backreacted solutions for the backgrounds can then be computed as in \autoref{fig:softIterations} which shows the results of the iteration process on the background fields (metric fields included). These are AdS-AdS domain wall solutions. \autoref{fig:softIterations} shows the fermionic stress tensor components \eqref{thermoQuantities} computed through the backreaction, vanishing in the UV and in the IR.}}

As a last remark, equation \eqref{eq:IRSchroAdS4} gives us a simple way to view the effect of the chiral-breaking flat potential. As has been pointed out in \cite{Faulkner:2009,Gubser:2009dt},
%[ALSO GUBSER \nc{Done}]
in \ads with constant electrostatic potential where $\lambda = 0$, the potential is deconfining for modes with $|\omega_\ir| > |k_\ir|$ and confining for modes such that $|\omega_\ir| < |k_\ir|$. 
% This means that pure \ads geometry would be confining for modes near $k_F$, but would fail to keep the low-momentum mode from propagating. 
The addition of a \textit{flat} profile means that now modes with $|k_\ir| \leq |\omega_\ir| < \sqrt{k_\ir^2 + \lambda_\ir^2}$, which previously were not bound states, also become confined. This allows the existence of a window $\omega_-(k)< \omega < \omega_+(k)$, with $\omega_\pm(k) \equiv q A_t(z_\ir) \pm \sqrt{k_\ir^2 + \lambda_\ir^2}$ where a discrete set of (gapped) modes  can be populated.
% Therefore $\sqrt{k_\ir^2 + \lambda_\ir^2}$ is the effective energy cutoff induced by the regulated infrared geometry.
% Section 2.5
% \subsection{Backreaction and convergence}
% Section 2.6
% \subsection{Removing the regulator}
% Section 2.7
% \subsection{Linear band}
% \label{sec:linearBand}
% \input{sections/swqes_linband}

% Section 3
\section{Regulated Quantum Electron Star: thermodynamics and spectrum}
\label{sec:dualqft}
\begin{figure}[t!]
    \hspace*{-.3in}
    \includegraphics[scale=0.8]{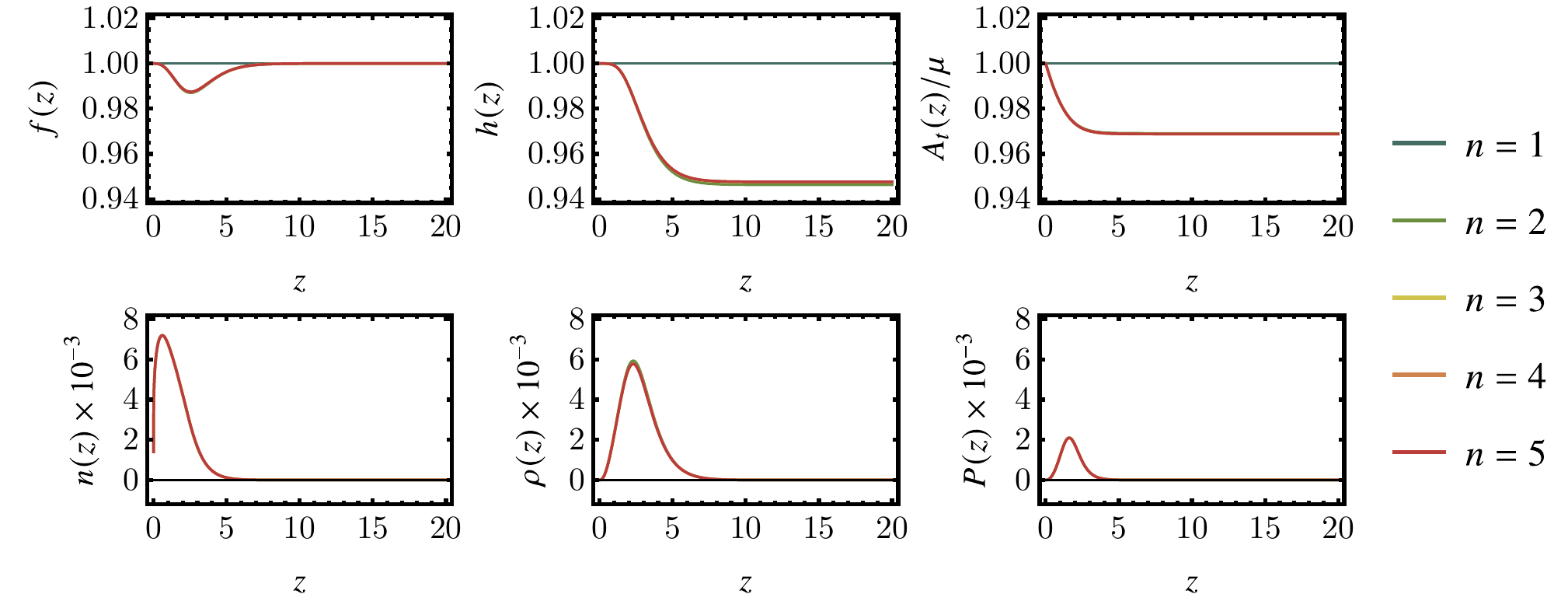}
    \caption{Iterative backreactions on the background fields ($f(z)$, $h(z)$, $A_t(z)$) and their associated currents $(n(z), \rho(z), P(z))$ with the same parameters as in \autoref{fig:softConv}. In total 5 iterations are performed, denoted by the color scale from green (first iteration) to red (last iteration). For these values $\{m,\mu q, \lambda\}=\{0.1, 0.9, 1\}$ only the first iteration differs significantly from the final solution, and the other curves are visually barely distinguishable from each other; for higher $q$ convergence rapidly becomes slower.}
    \label{fig:softIterations}
\end{figure}

Now that the problem is well-posed, we can follow the algorithm in \autoref{fig:algo} and construct a fully backreacted regulator-confined $T=0$ quantum electron star. Choosing the chirality-breaking flat regulator the resulting solution is shown in \autoref{fig:softIterations}. This is by construction an \ads-to-\ads domain wall solution. Just like the analogous domain wall solutions for the holographic superconductor \cite{Gubser:2008wz,GubserNellore,HorowitzRob}, it has a UV \ads and an IR \ads with the \emph{same} radius but different effective speed of light. This can be checked by considering the diffeomorphism-invariant ratios $v_{\ir}/v_{\uv}$ and $L_{\ir}/L_{\uv}$ which are equal to
\be
    \dfrac{L_\ir}{L_\uv} = \sqrt{\dfrac{R(z \to z_\uv)}{R(z \to z_\ir)}} = 1~, \quad \dfrac{v_\ir}{v_\uv} \equiv \dfrac{v(z \to z_\ir)}{v(z \to z_\uv)} =  \sqrt{\dfrac{h(z \to z_\ir)}{h(z \to z_\uv)}} < 1 \text{ in our solution}~.
\ee
Here $R(z)$ is the Ricci scalar and $v(z) = \sqrt{h(z)}$ is deduced from the null vector $\dfrac{\mathrm{d}}{\mathrm{d}t} X^\mu (z)$ where $X^\mu(z) \equiv \{ t, 0, v(z) t, 0 \}$ is a $x$-directed trajectory. Therefore, our solution obeys the $c$-theorem since the effective speed of light in the dual field theory is lower in the IR than in the UV, as discussed in detail in \cite{Gubser:2008wz}.
% \\
% However, as discussed in detail in \cite{Gubser:2008wz}, such a solution still obeys the $c$-theorem if the speed of light in the dual field theory is lower in the IR than in the UV. The speed of light is given by the coordinate speed $dx/dt$, corresponding to a null vector $(1,\sqrt{fh},0,0)$ so that $dx/dt=\sqrt{fh}$. As we can see in \autoref{fig:softIterations}, the function $h(z)$ asymptotes in the IR to some constant $0<h_0<1$, while $f(z)$ asymptotes to unity. Therefore the speed of light equals $0<\sqrt{h_0}<1$ as it should be.

In accordance with our discussion in the Introduction, the chemical potential is chosen such that only the lowest radial mode of the fermionic wavefunction is occupied.  The associated matter content shows that a localized distribution of fermions in the mid-infrared region 
% and the boundary 
is characterized by a stable finite density of fermions with total charge \mbox{$Q = - A_t^\prime(z \to 0)$}.

\begin{figure}[h!]
    \centering
    \includegraphics[scale=0.8]{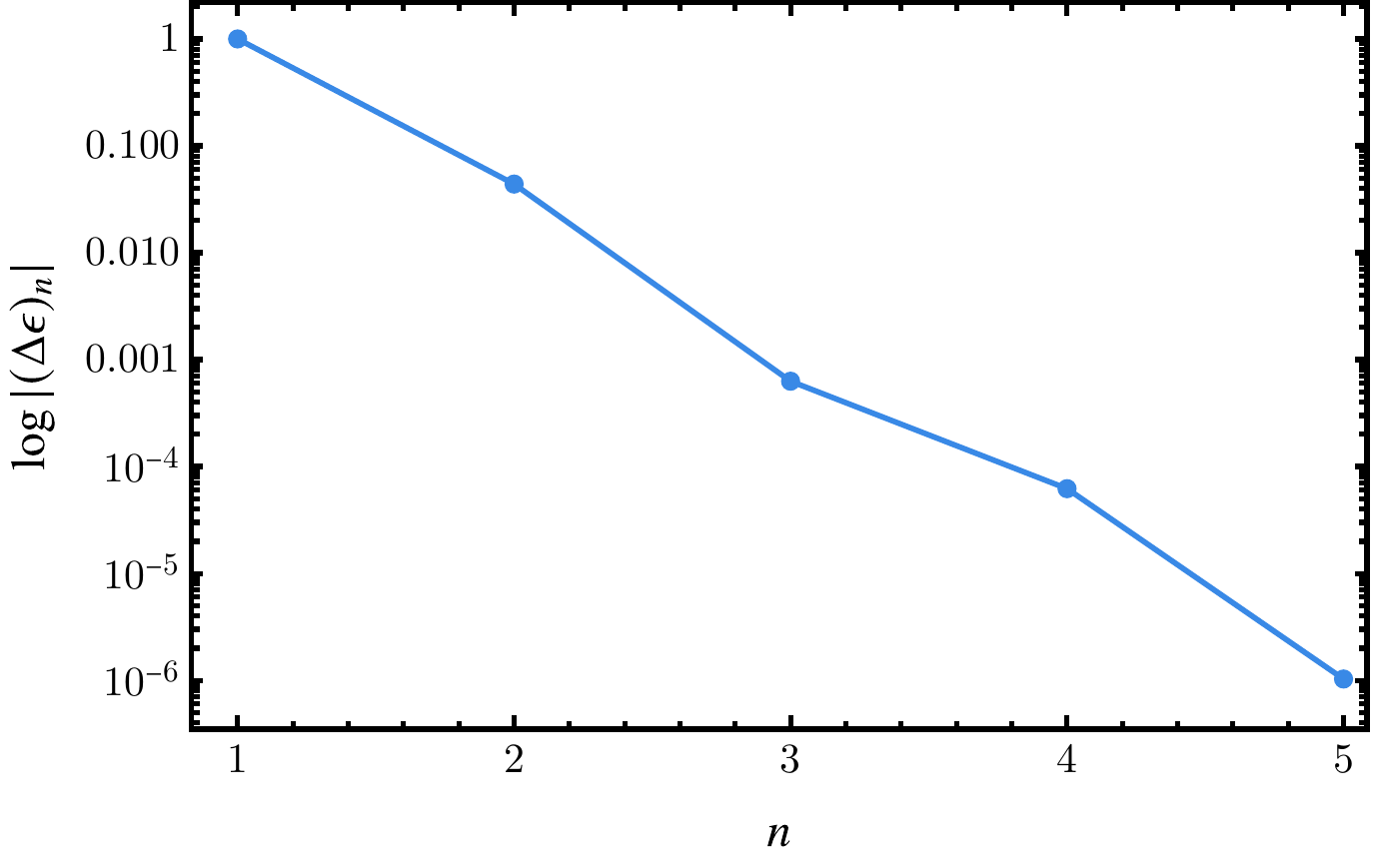}
    \caption{Convergence in terms of the logarithm of the difference in the IR between the $n$-th and $n+1$-st iteration $\log\vert(\Delta\epsilon)_n\vert$ for a rQES with  $\{m,\mu q, \lambda \} = \{ 0.1, 0.9, 1 \}$. The convergence is exponentially fast and the agreement is very good already around the 3\textsuperscript{rd} iteration.}
    \label{fig:softConv}
\end{figure}

With the chirality-breaking flat potential the convergence is in fact quite fast at low density. The Hartree algorithm provides
 a discrete sequence of fields $(f^{(n)}, h^{(n)}, A_t^{(n)})$ as we iterate from $n=1,2,\ldots$. We can introduce a criterion for the convergence of the solution using the IR parameters used for shooting
\begin{equation}
    \epsilon_n = \sqrt{f^{(n)}(z_\ir)^2 + A^{(n)}_t(z_\ir)^2 + h^{(n)}(z_\ir)^2}~,
\end{equation}
Convergence is obtained if $(\Delta\epsilon)_n \equiv \epsilon_{n} - \epsilon_{n-1} \xrightarrow{n \to \infty} 0$.
For a small occupation number/charge \autoref{fig:softConv} shows that the solution already 
stabilizes after three iterations; for large occupation numbers the convergence rapidly becomes much slower. We have checked that the solution is not sensitive to the choice of the numerical cutoffs $\{z_{\uv}, z_{\ir}\}$.

% \begin{figure}[H]
%     \centering
%     \includegraphics[scale=1]{figs/ffSW0.pdf}
%     \includegraphics[scale=1]{figs/hhSW0.pdf}
%     \includegraphics[scale=1]{figs/gaugeSW0.pdf}
%     \caption{Iterative backreactions on the background fields ($f(z)$, $h(z)$,$A_t(z)$ with $m = 0.1$, $q = 1$, $z_{\text{IR}} = 20$, $z_0 = 2$, $\mu = 0.8$, $\lambda = 1.5$, $\alpha = 4$. The bulk is asymptotically AdS in the UV and in the IR but with
%     different scalings.}
%     \label{fig:softIterations}
% \end{figure}

% \begin{figure}[H]
%     \centering
%         \includegraphics[scale=1]{figs/rhoFW0.pdf}
%         \includegraphics[scale=1]{figs/plonFW0.pdf}
%         \includegraphics[scale=1]{figs/densityFW0.pdf}
%     \caption{Thermodynamic fields ($n(z)$, $\rho(z)$ and $P(z)$) with $m = 0.1$, $q = 1$, $z_{\text{IR}} = 20$, $z_0 = 2$, $\mu = 0.8$, $\lambda = 1.5$, $\alpha = 4$. The bulk is asymptotically AdS in the UV and in the IR but with different scalings.}
%     \label{fig:softCurrents}
% \end{figure}

% Section 3.1
\subsection{Thermodynamics}
For a large $q/m$ ratio we expect that the quantum electron star at a given chemical potential $\mu$ is the thermodynamically preferred solution over the extremal \rn solution.
In order to study the thermodynamics of the regulated quantum electron star, we need to compute its free energy. It consists of two parts. There is a direct saddle point contribution from the regularized Euclidean action:
\begin{equation}
	S_E = \int \dd^4 x \sqrt{g_E} \left[ \dfrac{1}{2\kappa^2} \left( R + 6 \right) - \dfrac{1}{4} F^2 \right] + \oint_{z=\epsilon} \dd^3 x \sqrt{h} (-2 K + 2 \gamma)~,
\end{equation}
where $g_E$ is the Euclidean metric, $h$ is the induced metric on a hypersurface normal to a radial ($z$) slice, pointing outwards, $K$ is the trace of the extrinsic curvature and $\gamma = 2$ is required to make the AdS free energy vanish. 
The imaginary time at temperature $T$ is compactified with the radius $\beta=1/T$, the integral in the $x$--$y$ plane produces the (infinite) volume $\mathrm{Vol}_2$, and the radial integration is performed to some UV cutoff $\epsilon$,
yielding
\begin{equation}
	S_E = \beta \mathrm{Vol}_2 \int \dd z \sqrt{g_E} \left[ \dfrac{1}{2\kappa^2} \left( R + 6 \right) - \dfrac{1}{4} F^2 \right] + \beta \mathrm{Vol}_2 \sqrt{h(\epsilon)} (-2 K(\epsilon) + 2 \gamma)~.
\end{equation}
This accounts for the contribution of the bosonic fields. The Dirac action vanishes on-shell and therefore does not contribute to this part. It does have a one-loop contribution to
the free energy density 
\begin{equation}
	f \equiv \dfrac{S_E}{\beta \mathrm{Vol}_2} + f_{\mathrm{Dirac}}~.
\end{equation}
Here $f_{\mathrm{Dirac}}$ represents the fermionic contribution. Following \cite{Sachdev,Denef:2009yy,Iizuka:2011hg,Hashimoto:2012ti}, at $T=0$ we can simply
%consider the micro-canonical internal energy obtained by
sum the energies along the filled band of fermions (above the Dirac sea). This is the internal energy shifted by the chemical potential. For our normal modes, this leads to the expression
\begin{equation}
	f_{\mathrm{Dirac}} = \sum_{\ell} \int \dfrac{k \dd k}{2 \pi}~ \Theta(- E_{\ell}(k)){\Theta(E_{\ell}(k)-\mu q)} E_{\ell}(k) \nonumber
	= \int \dfrac{k \dd k}{2 \pi}~ \Theta(- E_{1}(k)) E_{1}(k)
	%< 0~.
\end{equation}
where in the last line we have made explicit that we choose our chemical potential such that only states of the lowest electronic radial mode $E_{\ell = 1}$ will be occupied. One must first choose the potential strength $\lambda$ such that the \sch potential supports at least one normalizable mode. At the same time, it is only these normalizable modes that can be populated. If there is only one band in the window of existence of normalizable modes $[\omega_-(k), \omega_+(k)]$, i.e.,$E_{\ell = 1}(k) < \omega_+(k) < E_{\ell = 2}(k)$, then increasing the chemical potential beyond that upper limit will not populate further normalizable modes. Our rQES is in this sense not plagued by the usual large-$N$ Fermi surfaces artifact.

\begin{figure}[t!]
	\centering
	\includegraphics{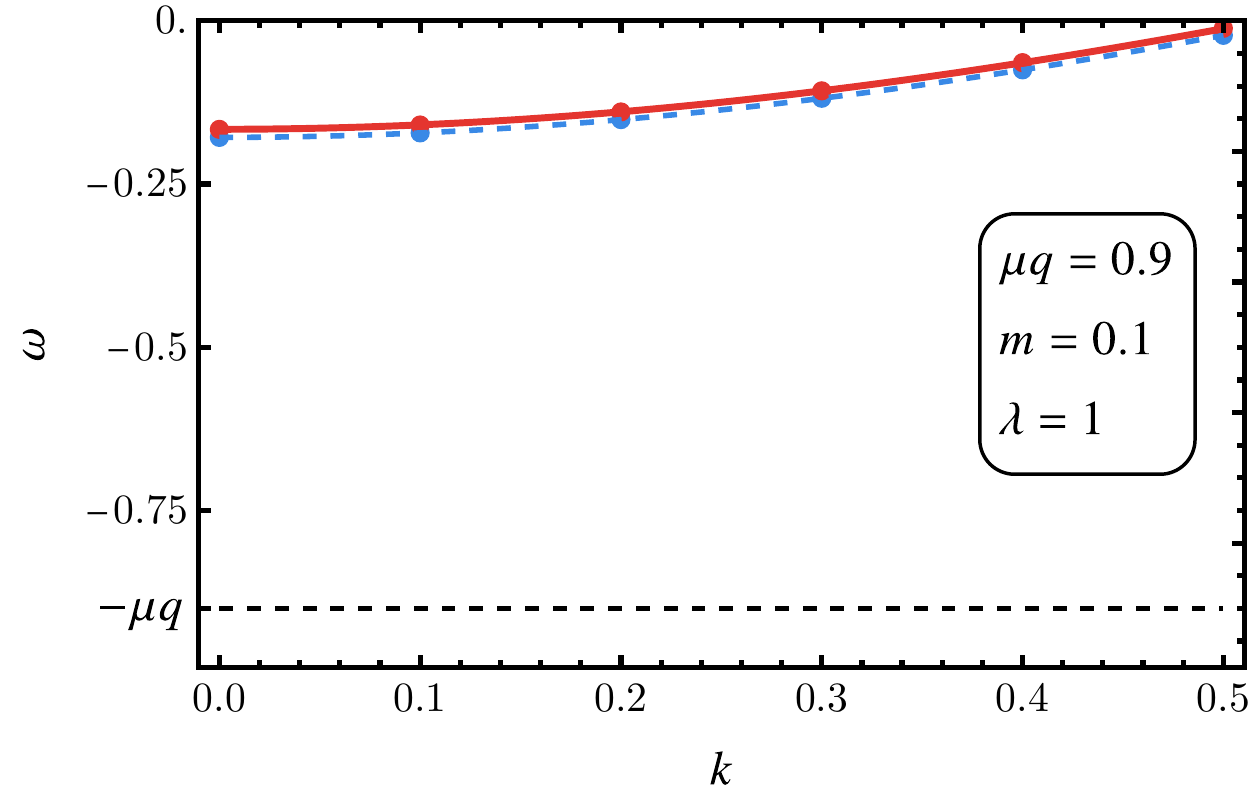}
	\caption{First electronic band for $\{m,\mu q, \lambda \} = \{ 0.1, 0.9, 1 \}$, for the \ads background with constant electrostatic potential (blue) and the backreacted solution (red). The lines are a fit to the form \eqref{eq:symbolicBand}. }
	\label{fig:fitBand}
\end{figure}

It is furthermore quite easy to show that both before and after accounting for backreaction the band structure follows a similar form as in pure \ads \cite{Sachdev}
\begin{equation}
    \label{eq:symbolicBand}
	E_\ell(k) = - E_0 + \sqrt{k^2 + k_0^2}~,
\end{equation}
where $k_F \equiv \sqrt{E_0^2 - k_0^2}$ and the parameters $E_0$, $k_0$ are most easily found by fitting from the numerical dispersion curves, as in \autoref{fig:fitBand}.

% \begin{figure}[t!]
% 	\centering
% 	\includegraphics{figures/fitBands.pdf}
% 	\caption{Band structure for $\mu q \in \{0.75, 0.825, 0.9, 0.975, 1.05\}$ at fixed $m = 0.2, \lambda = 1$. For these small values of $\mu q$, we still have the approximate relation $E_0 \simeq \mu q$ which becomes exact in pure \ads. The dashed lines are a fit to the form \eqref{eq:symbolicBand}. \ks{[IS THIS IN THE BACKREACTED STAR OR PURE ADS. IF USEFUL PLEASE SHOW CONVERGENCE HERE AS WELL]}}
% 	\label{fig:fitBand}
% \end{figure}

Note that $f_{\text{Dirac}}$ is negative semi-definite. This does not mean, however, that the occupied state is automatically thermodynamically preferred. The backreaction also changes the bosonic saddle point contribution compared to its original AdS$_4$ value $f(\text{AdS}_4)=0$. 
Adding both contributions we compare to the RN free energy
\begin{equation}
	%f\left(\mathrm{AdS}\right) = 0~, \quad 
	f\left(\mathrm{RN}\right) = - \dfrac{4 + z_h^2 \mu^2}{4 z_h^3} = - \dfrac{\mu^3}{6 \sqrt{3}} \text{ at } T = 0~.
\end{equation}
Because the regulator does not act on the background sector, the \rn free energy is unaffected by it.

\begin{figure}[t!]
	\centering
	\includegraphics[width=\textwidth]{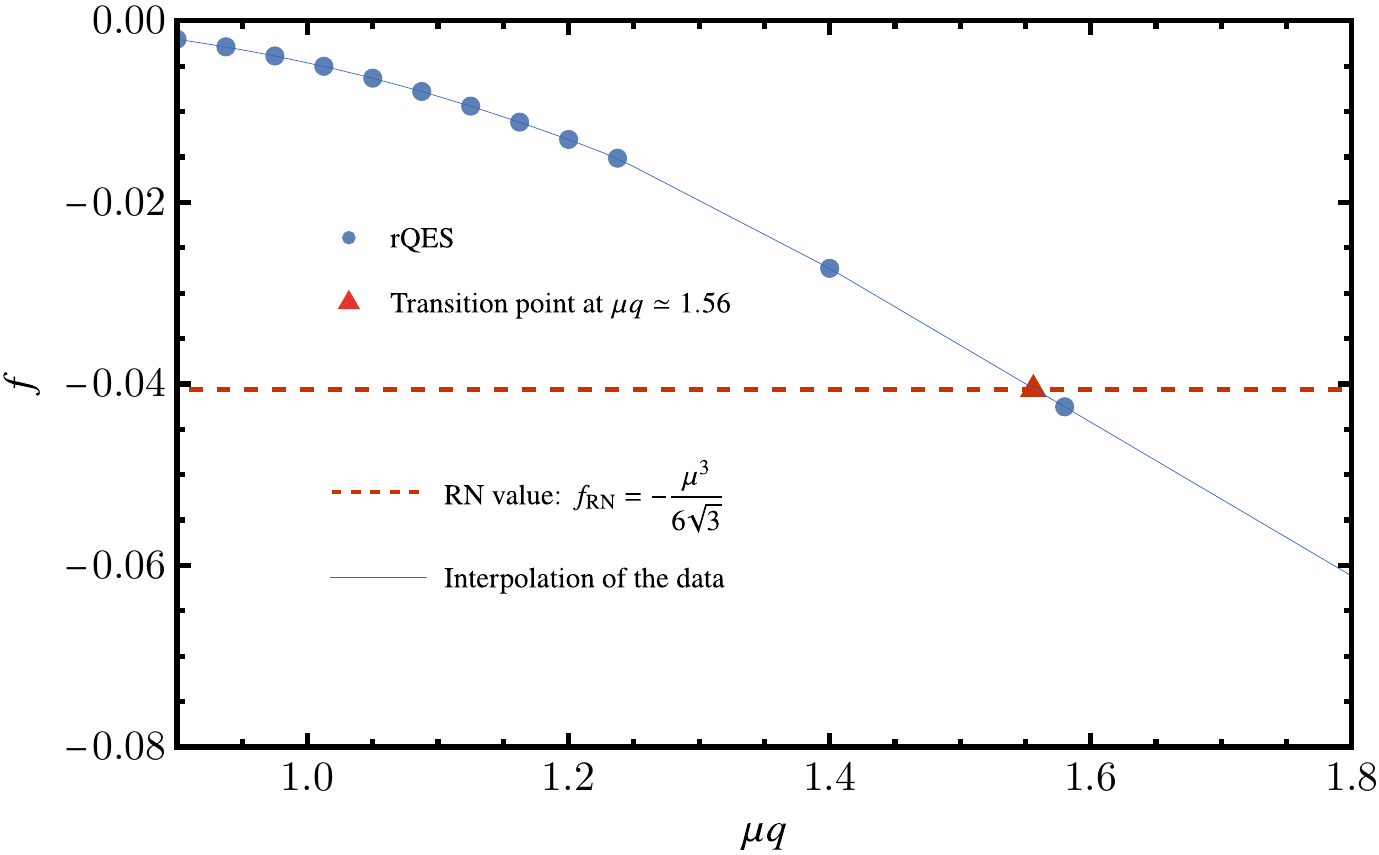}
	\caption{Plot of the free energy density for rQES at $\{\mu, m, \lambda\} = \{0.75, 0.1, 1\}$ as a function of the fermionic charge $\mu q$ (blue dots) and the reference RN black hole free energy (red dashed line); the thin blue
	line and the red triangle are to guide the eye to the transition point. Since RN has no fermions its free energy curve is flat, i.e. does not depend on the fermion charge. The first-order phase transition from RN to rQES happens at the intersection of the two
	lines. Since the calculations for larger $\mu q$ values are costly, we only compute two points for $\mu q>1.5$ and interpolate.
	%\ks{[MAKE $f_\rn=-\mu^3/6\sqrt{3}$. Can we also do $\mu q=1.4$ and $1.6$?]}
	}
	\label{fig:freeEnergy}
\end{figure}

% We have seen from the solution of our regulated quantum electron star that as we increase the value of the coupling $q$, more fermionic modes become occupied. This in turn means that their backreaction on the geometry increases and
% the background deviates more and more from the AdS geometry. This deformation of the background as well as the internal contribution of the fermions lowers the free energy until a critical value $q_c$ is reached for which $f = f_{\mathrm{RN}}$.

\autoref{fig:freeEnergy} shows the free energy of the rQES as a function of the charge $\mu q$ for a fixed mass $m$ and confining potential strength $\lambda$. As $q$ increases, the rQES grows, so we need to compute more and more modes. This becomes more and more time consuming. By constructing an interpolating curve based on low $q$ rQES solutions (using the points until $\mu q \simeq 1.2$), we can estimate where the solution becomes thermodynamically preferred and verify this with a fewer number of large $q$ datapoints ($\mu q = 1.4$ and $\mu q = 1.58$).  
% For that reason we only do
% a few points for large $q$ values and construct an interpolating curve. Even so, it is clear that 
We see that at $\mu q = \mu q_c \simeq 1.56$, the rQES becomes thermodynamically preferable over the RN background.

\begin{figure}[t!]
     \centering
     \begin{subfigure}[c]{0.44\textwidth}
         \centering
         \includegraphics[width=\textwidth]{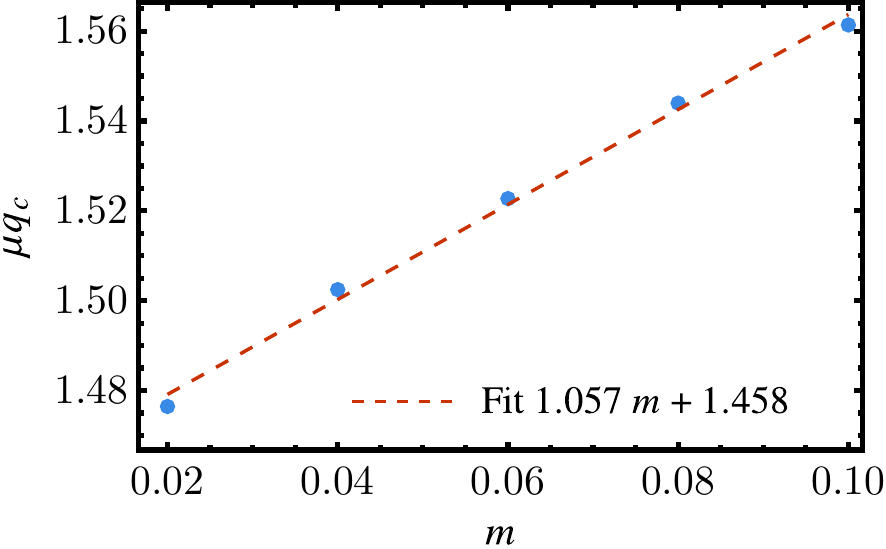}
         \caption{}
         \label{fig:fitTransitionPoint}
     \end{subfigure}
     %\hfill
     ~~
     \begin{subfigure}[c]{0.44\textwidth}
         \centering
         \includegraphics[width=\textwidth]{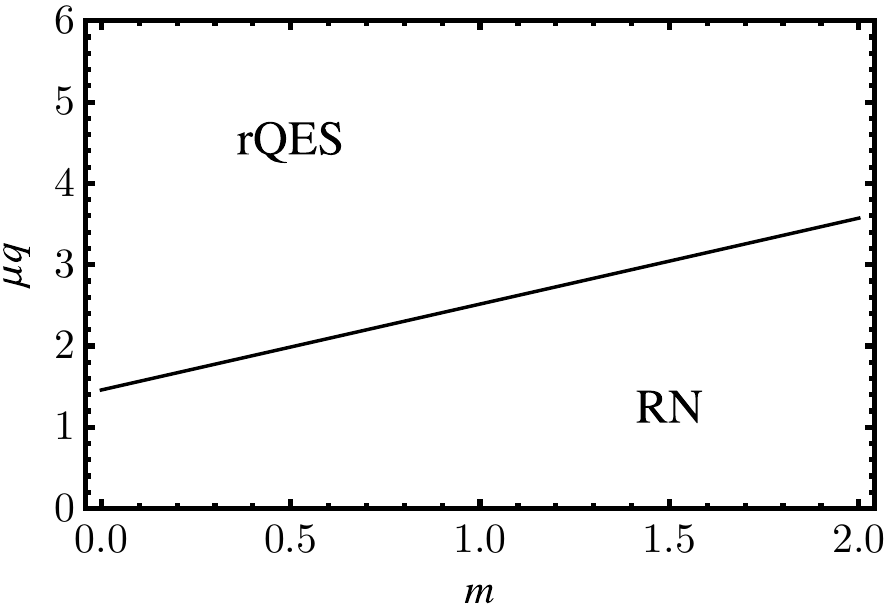}
         \caption{}
         \label{fig:phaseSpacerQES}
     \end{subfigure}
     \caption{
     	(\subref{fig:fitTransitionPoint}) Transition point $\mu q_c$ as a function of $m$ and its linear fit, for $\lambda = 1$.
     	%\ks{[CAN YOU MAKE THE BLUE DATA POINTS BIGGER]} %Linear dependence is obvious, although we do not understand it analytically.
     	(\subref{fig:phaseSpacerQES}) Sketch of the phase diagram of the rQES. The black line indicates a first order transition between the regulated \rn and the rQES, occuring when their free energies cross.
     }
\end{figure}

In \autoref{fig:fitTransitionPoint}, we show that this transition point evolves linearly with the fermion mass $m$ for fixed $q$ and $\lambda$. Based on this finding, we can sketch a thermodynamic phase diagram for our model in \autoref{fig:phaseSpacerQES}. The critical charge satisfies
an approximate relation $q_c(m;\lambda)\approx c_0(\lambda) + c_1(\lambda) \frac{m}{\mu}$ with $c_0$ and $c_1$ dependent on $\lambda$. 
%\ks{DO WE KNOW HOW IT BEHAVES AS A FUNCTION OF $\lambda$? ]}. 
It is tempting to compare this to the confounding phase diagram based on RN holography alone. For pure RN holography it is surmised  \cite{Liu:2011} that the superradiant instability of the RN black hole toward an electron star (seen in the spectrum as log-periodic oscillations)
sets in at $q=\sqrt{3}m$. This should correspond to the limit $\lambda \rightarrow 0$. As $\lambda$ decreases we therefore expect the phase-boundary to pivot anti-clockwise. This comparison should be done with care, because the smaller $\lambda$ becomes, the harder it is to observe bands that can be occupied --- see the section on removing the regulator below. Another way to see this is that the
effective \sch potential in the extremal RN black hole for $\omega=k=0$ (the onset of instability) has no linear term in $m$: $V_{\text{RN}}\sim -4q^2+2m^2$.
%with some numerical factors $c_1,c_2$, 
%Hence we cannot say anything
%about the $m$-dependence from the simplest approximation; for sure there is no reason that it should be the same as for extremal RN.
Hence we cannot extrapolate freely to $\lambda=0$.

% Section 3.2
\subsection{Spectrum of the rQES}
%\subsubsection{Derivation of the fermionic retarded Green's function}
To confirm our results, we consider the fermionic spectral function on rQES backgrounds.
% the true and tried way to get some insight into the states and excitations of the system. 
% In order to do this, we follow the well-established holographic way and add
% a probe Dirac fermion coupled to the same scalar regulator as the background Dirac fermion. 
As a reminder, the spectral function is defined as the trace of the imaginary part of the retarded propagator: $A(\omega,k)=\Im\Tr G_R(\omega,k)$. In holography the type of propagator is defined by the boundary conditions in the interior. Therefore the only difference with computing the normalizable Dirac solutions is the  choice of appropriate boundary conditions.

Considering that we have an emergent \ads geometry in the IR, we can use the known prescription for infalling boundary conditions in pure AdS, i.e. the presence of a Poincar\'{e} horizon \cite{Iqbal:2009fd}. %Therefore, the boundary conditions in the IR 
Accounting for the confining potential, these are
\begin{equation}
	\label{eq:IRBCInf}
	\psi_1(z \to \infty) = \begin{cases}
		e^{- z \sqrt{\bm{k}_\ir^2}}~, & \text{ if } \omega_\ir^2 < k_\ir^2 + \lambda_\ir^2~, \\
		e^{i z \sqrt{-\bm{k}_\ir^2}}~, & \text{ if } \operatorname{Re}[\omega_\ir] > \sqrt{k_\ir^2 + \lambda_\ir^2}~,\\
		e^{- i z \sqrt{-\bm{k}_\ir^2}}~, & \text{ if } \operatorname{Re}[\omega_\ir] < - \sqrt{k_\ir^2 + \lambda_\ir^2}~,\\
	\end{cases}
\end{equation}
where $\omega_\ir, k_\ir, \lambda_\ir$ were defined by \eqref{eq:IRSchroAdS4}, $\bm{k}_\ir = (\omega_\ir, \sqrt{k_\ir^2 + \lambda_\ir^2}, 0)$ and $\bm{k}_\ir^2 = - \omega_\ir^2 + k_\ir^2 + \lambda_\ir^2 = V_\ir$. As we saw with the normal modes, the IR boundary condition for $\psi_2$ can be obtained using the Dirac equation and the boundary condition for $\psi_1$. After imposing these boundary conditions, the retarded propagator is then computed  as 
\begin{equation}
	G_R(\omega, k) = B/A = \lim_{z \to 0} z^{-2 m} \dfrac{\psi_1(z)}{\psi_2(z)}~,
\end{equation}
where $A$ and $B$ are the coefficients in the UV expansion of the spinor \eqref{bndpsi}. 

Inside the gap ($\omega_{\ir}^2<k_{\ir}^2+\lambda_{\ir}^2$) the IR boundary conditions are the same for the probe fermions as for the
bulk normalizable modes -- the wavefunction should fall off for $z\to\infty$, which yields $A = 0$ for the normal mode frequencies $\omega = E_\ell(k)$. Therefore, the propagator will present a pole along the bands of the background. Moreover,
since the fermionic wavefunctions and thus also the Green's functions are real inside the domain where bound states exist, the spectral function will vanish there. Thus, we expect to see $\Im G_R(\omega, k) = 0$ for
$\omega\in\left[\omega_-\left(k\right),\omega_+\left(k\right)\right]$, except when $\omega = E_\ell(k)$ where a pole should appear.

%\subsubsection{Numerical results}

%Now we are ready to discuss the spectra as found from the numerics. 

\begin{figure}[t!]
	\centering
	\includegraphics[scale=1]{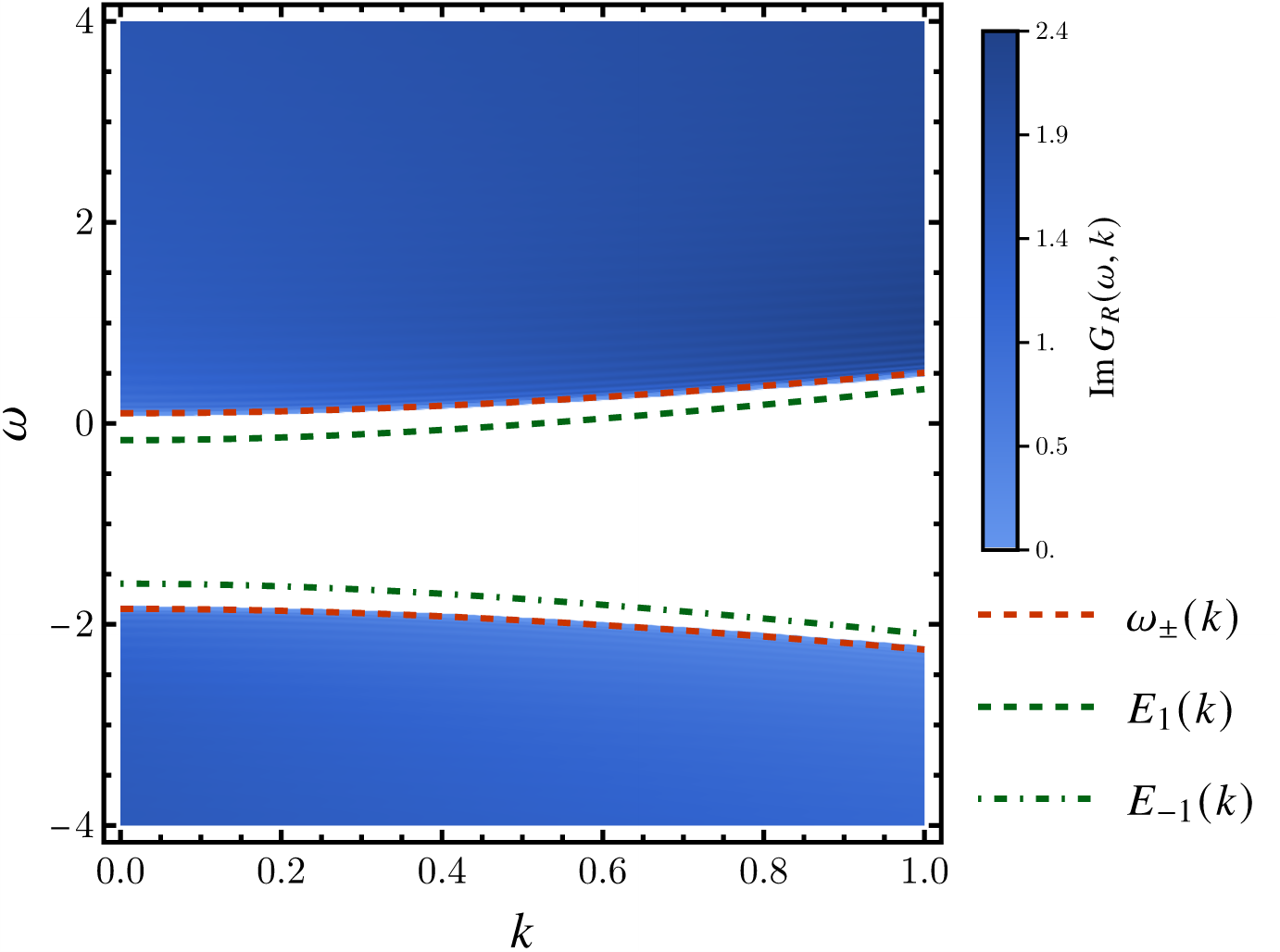}
	\caption{Spectral function $\Im G_R(\omega,k)$ for $\{m, \mu q, \lambda\} = \{ 0.1, 0.9, 1 \}$. The gap appears in white and is well delimited by $\omega_\pm(k)$ (red dashed lines). The normal mode bands have been superimposed to show the infinitely long-lived
	modes, see \autoref{fig:plotAllSpectral}. Outside the gap, there is no particle (normal mode) but a continuum shaped by the remnant of the UV conformal branch cuts. Since the regulator and the chemical potential explicitly break conformality, we do not reproduce the pure AdS Lorentz-invariant spectrum for any finite value of $\omega$ and $k$.}
	\label{fig:fullSweep}
\end{figure}

\begin{figure}[tbp]
     \centering
    \includegraphics{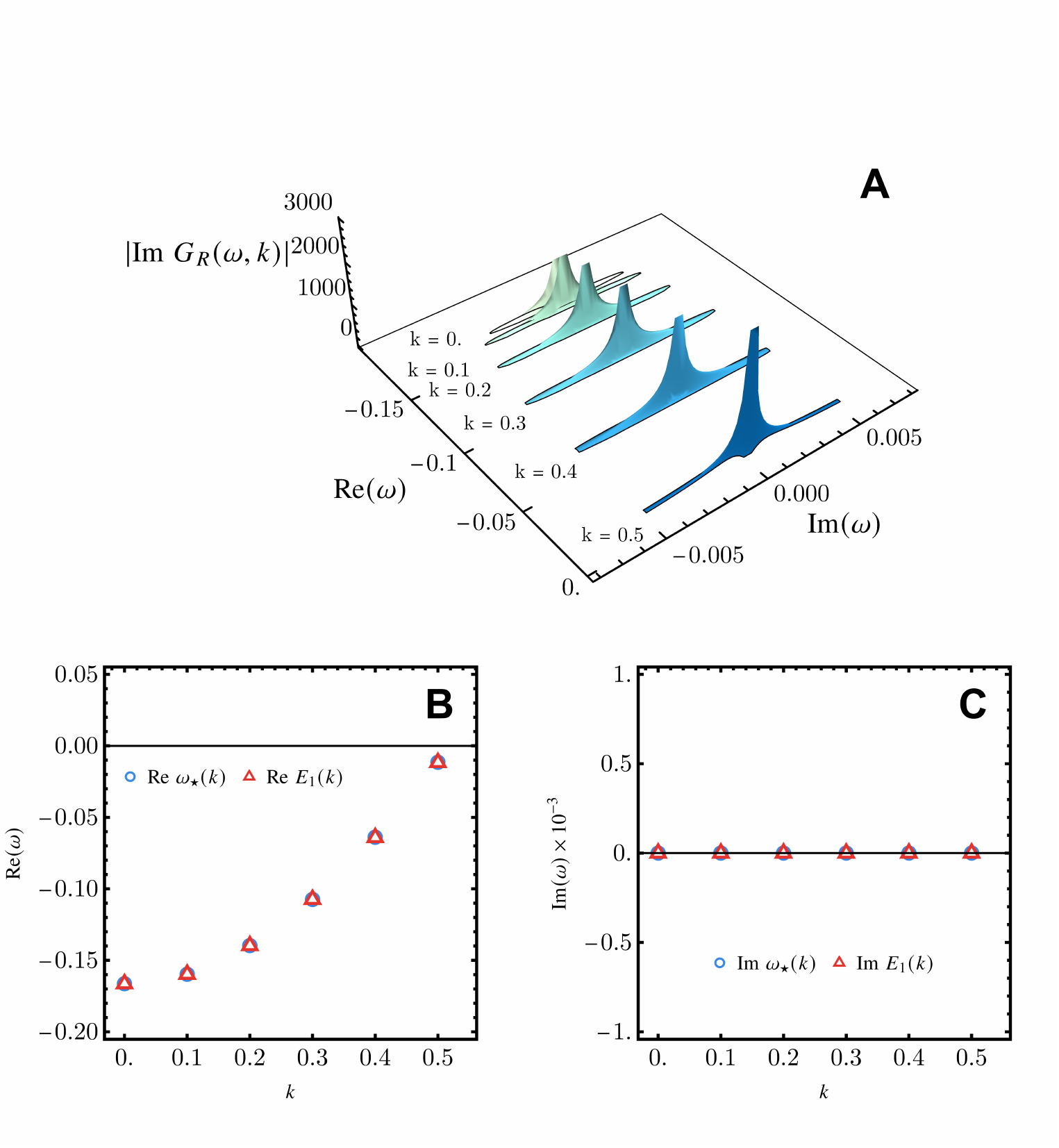}
     \caption{
     	(A) Absolute value of the fermionic spectral function for different values of momentum. The plot is cropped for values below $100$ to highlight the quasiparticle peaks.
        (B and C) Comparison of the poles in the spectrum (blue circles), identified in (A), to the first electron band of the background (red triangles). The real parts (B) of both sets agree
         perfectly; the imaginary parts (C) are both zero to high accuracy. All this data is computed for  $\{m, \mu q, \lambda\} = \{ 0.1, 0.9, 1 \}$.
         %\ks{[CAN WE REFORMAT THESE FIGURES SO THEY ARE ALL THE SAME SIZE?]}
     }
     \label{fig:plotAllSpectral}
\end{figure}
This general structure of the spectral function including the gap for
$\omega_- \leq \omega \leq \omega_+$ can be seen in \autoref{fig:fullSweep}. The data here and in the remainder of this section is computed for $\{\mu, q, m, \lambda\} = \{3/4, 1.2, 1/10, 1\}$. Inside the gap (white area), the spectral weight of excitations is indeed zero to numerical accuracy {except} at the positions of the normal modes of the background fermions. The latter are computed
%them 
directly from the solution of the background Dirac equation (green lines in \autoref{fig:fullSweep}),
%(blue
%and yellow dots in \autoref{fig:fullSweep} \ks{[]THERE ARE NO BLUE/YELLOW DOTS]}), 
%however 
as they cannot be seen numerically in the spectral function 
%for \emph{real} $\omega$ 
because they are infinitely long-living modes which show in the spectrum as Dirac delta peaks. Being infinitely
narrow on the real axis, they can only be detected in the complex-$\omega$ plane. Representing schematically the normal mode located at $\omega_\star$ by Im$G(\omega=\text{Re}(\omega)) = Z\delta(\omega-\omega_\star)$ where $Z$ is the peak weight (wavefunction
renormalization), we have, for complex $\omega$:
\begin{equation}
	\label{eq:poleSpectralFunction}
	\Im G_R(\omega, k) = -Z \dfrac{\Im\omega-\Im\omega_\star}{\left(\Re\omega-\Re\omega_\star\right)^2+\left(\Im\omega-\Im\omega_\star\right)^2}~.
\end{equation}
When $\Re\omega = \Re\omega_\star$, this simplifies to
\begin{equation}
\label{eq:poleSpectralSimple}
	\Im G_R(\omega, k) = -\dfrac{Z}{\Im \omega - \Im\omega_\star}~.
\end{equation}
We check this picture against the numerics first in \autoref{fig:plotAllSpectral} (A), where the absolute value of the spectral function in complex frequency plane shows the typical structure of a string of poles (for various momentum values) lying on the real
axis. The relation (\ref{eq:poleSpectralSimple}) is then used to identify the dispersion relation of the pole $\omega_\star(k)$ by fitting $\Im G_R(\omega, k)$. We find, with no big surprise, a perfect agreement with the normal mode excitations $E_1(k)$ corresponding to the first electron band, as seen in \autoref{fig:plotAllSpectral} (C) and (D). A similar picture is
found for the first hole band $E_{-1}(k)$ and this yields the spectrum inside the gap, plotted in \autoref{fig:fullSweep}.

\bigskip

\begin{figure}[]
	\centering
	\includegraphics[width=\textwidth]{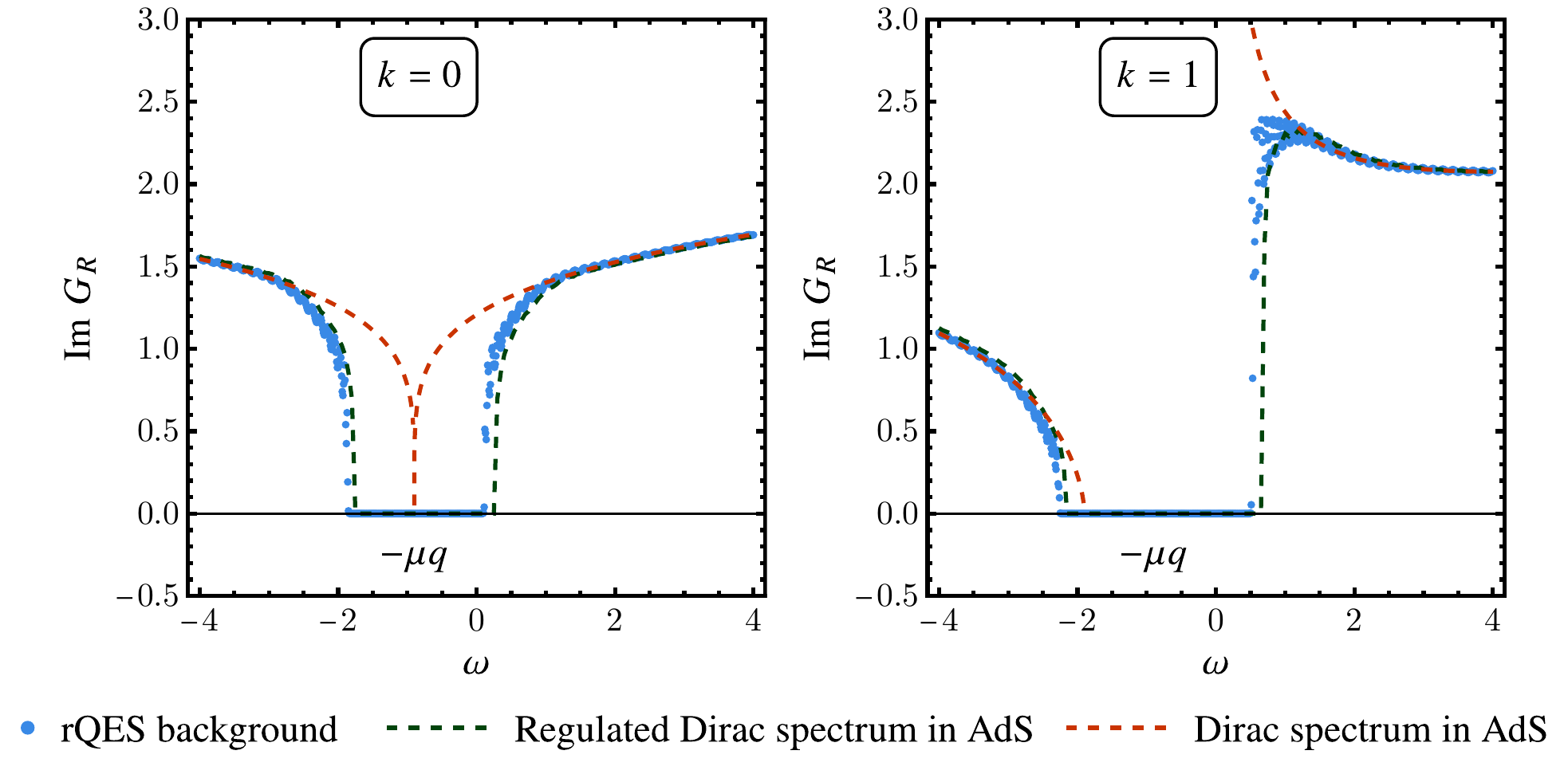}
	\caption{(Confined) Dirac spectral function (blue points) in the rQES background for $k = 0$ (left) and $k=1$ (right), compared with the standard/unconfined (red dashed line) and regulated/confined (green dashed line) Dirac spectral function in AdS with finite electrostatic potential.
	%\mc{[KS: RELABEL "Domain wall background" --> "rQES background" AND "Confined Dirac" ---> "Regulated Dirac"]}.
	%\ks{FIX LABELING Regulated AdS --> confined Dirac spectrum in AdS}}and $\{m, \lambda, \mu q\} = \{ 0.1, 1, 0.9 \}$
	}
	\label{fig:spectralFunctionVsAdS}
\end{figure}

%\ks{[I DO  NOT KNOW WHAT THIS EQUATION MEANS, DROP?] \nc{This equation means that outside the gap, the spectral function is that of a renormalized  regularized AdS}}
In \autoref{fig:spectralFunctionVsAdS} we compare the spectral function at finite $\mu$ for our regulated quantum electron star (blue data points) to the fermionic spectral function in a pure \ads background with finite chemical potential, either with (green line) and without (red line) regulation by the confining scalar. The comparison is given at $k = 0$ (left) and $k = 1$ (right). The Dirac spectrum in \ads is well-known \cite{Iqbal:2009fd}:
\begin{equation}
    G_R(\omega, k) = \begin{cases}
        \dfrac{2}{\omega^2 - k^2} \dfrac{\Gamma(1/2 - m)}{\Gamma(1/2 + m)} \left[-\dfrac{i}{2} \left( \omega^2 - k^2 \right) \right]^{2m + 1} \left[\omega \gamma^0 - k \gamma^1 \right] & \text{ if } \omega > k~,\\
        \dfrac{2}{\omega^2 - k^2} \dfrac{\Gamma(1/2 - m)}{\Gamma(1/2 + m)} \left[\dfrac{i}{2} \left( \omega^2 - k^2 \right) \right]^{2m + 1} \left[ \omega \gamma^0 - k \gamma^1 \right] & \text{ if } \omega < -k~.\\
    \end{cases}
\end{equation}
It has a conformal branch-cut at $\omega = k$ and a gap for $\omega^2 < k^2$. For \ads with finite electrostatic potential, one merely needs to replace $\omega \to \omega + \mu q$ in the previous expression.
%\ks{[IS $\omega$ IN THE ABOVE EQUATION SECRETELY $\omega+\mu q$?]}.
Adding confining potential by turning on the chirality-breaking flat scalar widens the gap to $(\omega + \mu q)^2 < k^2 + \lambda^2$; in particular the gap is open also at $k = 0$. 
%and explicitly breaks the conformal symmetry leading to the disappearance of the branch cut. 
The rQES solution outside the gap exhibits qualitatively the same spectral function as that of the confined Dirac spectrum in pure \ads but for renormalized IR values $\omega_\ir, k_\ir, \lambda_\ir$ given in \eqref{eq:IRSchroAdS4}. It is important to emphasize that none of the modes in this continuum are normalizable and thus do not contribute when building the bulk rQES, even when $\mu q$ is large enough that $\omega_+(k) < 0$. This is guaranteed by our choice of UV boundary conditions.

% Section 4
\section{Towards a self-confining quantum electron star}
\label{sec:removingReg}
\subsection{Comparison to the holographic superconductor}

By construction the confinement in our setup gives an \ads-to-\ads solution. With the fully backreacted solution in hand we can also understand what the field theory dual describes. The confining regulator scale $\lambda$ gaps the field theory fermion spectral function. Considering then the RG flow from the IR emergent conformal field theory towards the UV, this means that as one increases the energy scale it takes a finite distance for occupiable fermion states to be encountered. This can also be seen in the band structure of \autoref{fig:fitBand}.
%[PUT $\omega=-\mu$ IN THERE].
At this scale the theory deforms away from the strict conformal theory up to the scale $\mu$ beyond which it is no longer energetically favorable to occupy more states. The flow up the RG then continues towards the UV \ads fixed point.

In the more usual flow from the UV to the IR this is not a natural RG trajectory. The generic IR will not be a non-trivial conformal field theory. Nevertheless, within holography such \ads-to-\ads domain walls are well-known. Especially in the search for the holographic dual of the holographic superconductor ground state, Horowitz and Roberts and independently Gubser and Rocha have found \ads-to-\ads domain walls (in some cases with logarithmic corrections) in a finite parameter range \cite{HorowitzRob,Gubser:2008wz}; the other solution found is the Lifshitz geometry. It was later understood that Lifshitz rather than an \ads IR is the generic holographic superconductor ground state \cite{Gubser:2008wz,GubserNellore}, but this is only seen with the inclusion of a stabilizing quartic potential. %Without this addition the RG-flow in that direction is blocked. 

In detail of course the solutions are different. The Horowitz-Roberts-Gubser-Rocha holographic superconductor ground states do not need an additional confining scalar. They can also be obtained classically without the need for a one-loop Hartree mean field. This is due to the fact that the bosonic field already couples quadratically to the electrostatic potential $A_t$. A fermion only couples linearly, but its one-loop contribution can couple at all orders. This is why for fermionic systems one needs to go to one-loop.

\subsection{Confinement in the rQES solution}

Given that the Horowitz-Roberts-Gubser-Rocha \ads-to-\ads solutions do not need a confining potential, and that the more generic holographic superconductor Lifshitz solutions are known, it is a natural question why we do not try to remove the soft-confining regulator alltogether. 
There was in fact a concerted effort to do so several years ago \cite{Hartnoll:es,Leiden:2011,Leiden:2013}, culminating in the QES model of \cite{McGreevy:2012,McGreevy:2013}. The latter two articles show in detail how the presence of the gap and the discretized spectrum are crucial to construct any type of quantum fermionic backreacted solution, i.e. where one or a small finite number of radial modes are occupied. Any attempt to remove the confining potential results in a uncontrolled continuum spectrum. 

It is precisely this insight that was the starting point for our confining potential. What we have furthermore shown, is that even then there are several severe technical hurdles to overcome to construct a converging fully backreacted confined quantum electron star solution. At the same time the general insight still holds.
Our infrared boundary conditions crucially depend on the coupling to the scalar $\Phi(z)$ to extend the domain of existence of normalizable modes of \ads all the way to $k = 0$. The parameter $\lambda$, as we previously noted, acts as a momentum shift in this domain such that a mode at $k = 0$ will behave as a mode at $k_{\mathrm{eff}} = \lambda$ and therefore normalizable modes with $|\omega + \mu q| < \lambda$ will be found. These can be populated and will condense in the bulk. Turning off the potential, even slowly, will invariably lead to a lack of normalizable modes at the lowest momenta and will bring us back to a situation similar to that of \ads.

\begin{figure}
     \centering
     \includegraphics{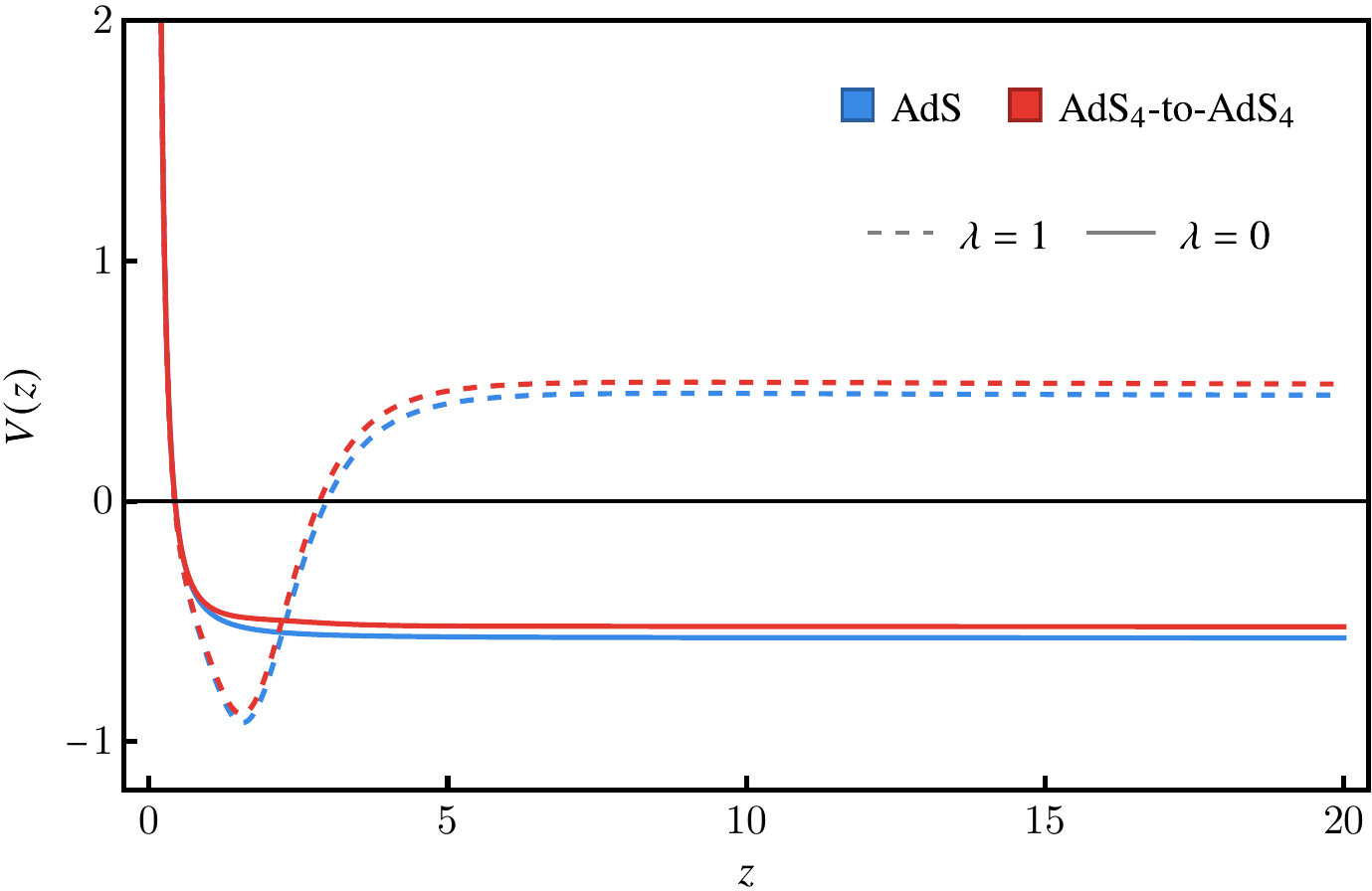}
	 \caption{Comparison of the \sch potential for the \ads (blue) and \ads-to-\ads (red) solutions with (dashed) and without (solid) regulator, for $\{m, \mu q, k, \omega\} = \{0.1, 1.05, 0, -0.027\}$.
	 %The curve in green is the \sch potential in pure \ads with IR renormalized parameters $\omega_\ir, %k_\ir$.
	 }
	 \label{fig:compareSchrodPot}
\end{figure}

One sliver of hope would be that the domain wall solution itself, after convergence, can support a well in the \sch potential such that a regulator is no longer necessary.
%\ks{\sout{We have therefore looked at such a question} (\autoref{fig:compareSchrodPot}) \sout{by comparing the \sch potential for a $k = 0$, $\omega = E_1(0)$ mode in the pure \ads background and the \ads-to-\ads domain wall solution.} [I BELIEVE YOU MEAN] }
We have therefore looked at this (\autoref{fig:compareSchrodPot}) by comparing the \sch potential for a $k = 0$, $\omega = E_1(0)$ mode in the confined quantum electron star \ads-to-\ads background with and without the confining potential. Without a potential, however, the \ads-to-\ads quantum electron star domain wall solution is not confining. We do see that $V_{\mathrm{domain\, wall}}(z \to \infty) > V_{\mathrm{AdS}}$ which means the wedge of existence of normalizable modes is indeed wider in the domain wall solution than in the \ads solution. Yet, the modes with sufficiently small momenta (including $k=0$) are always outside the wedge.

This therefore leads us to believe a true QES would not \emph{remove} the regulator but must \emph{incorporate} it into the model, i.e. make the scalar field a dynamical dilaton which couples to the Dirac fermion and drives the geometry from one fixed point to another.

% \begin{figure}
%      \centering
%      \begin{subfigure}[b]{0.48\textwidth}
%         \centering
% 	    \includegraphics{figures/comparisonSchrodPot.pdf}
% 	    \caption{Comparison of the \sch potential for the \ads (blue) and \ads-to-\ads (red) solutions with (dashed) and without (solid) regulator, for $\{m, \mu q, k, \omega\} = \{1/10, 1.05, 0, -0.027\}$. The curve in green is the \sch potential in pure \ads with IR renormalized parameters $\omega_\ir, k_\ir$.}
% 	    \label{fig:compareSchrodPot}
%      \end{subfigure}
%      \hfill
%      \begin{subfigure}[b]{0.48\textwidth}
%          \centering
% 	    \includegraphics{figures/IRPlot.pdf}
% 	    \caption{Comparison of the \sch potential for the \ads (blue) and \ads-to-\ads (red) solutions with (dashed) and without (solid) regulator, for $\{m, \mu q, k, \omega\} = \{1/10, 1.05, 0, -0.027\}$. The dotted curve is the \sch potential in pure \ads with IR renormalized parameters $\omega_\ir, k_\ir$.}
% 	    \label{fig:IRPlot}
%      \end{subfigure}
%         \caption{\ks{[I DO NOT UNDERSTAND THE POINT OF FIG(b). THIS IS JUST A ZOOMED IN VERSION OF FIG(a).]}}
%         \label{fig:three graphs}
% \end{figure}

% Section 6
\section{Discussion and conclusions}\label{sec:conc}
In this paper we have constructed a self-consistent model
% although somewhat artificial model 
of a single band confined holographic Fermi liquid. The crucial technical problem, the infrared divergence brought about by the fermionic wavefunctions, is solved by controlling it by hand. %rather drastic means; 
We
%cut off mercilessly 
control
the far infrared by the means of a scalar regulator, equivalent to a soft-confining potential. The confinement is drastic and 100\%: 
%and the price to pay is quite high: 
our regulated quantum electron star is dual to a gas of infinitely-long living particles with zero self-energy. In the limit where we compute, it is a single-band Fermi-gas rather than a Fermi-liquid.\footnote{This holds at zero 
temperature. At finite temperature a black hole horizon would form, causing inevitably some dissipation even in the presence of the confinement.} At higher energies, the spectrum switches 
%the infrared \ads leads 
to the featureless continuum inherited from the UV conformal field theory (though it 
%(which 
is 
not conformally invariant due to the presence of the confining potential). 

%On the other hand,
The regulated quantum electron star is the thermodynamically preferred solution over the \rn background for $\mu q/m > (\mu q/m)_{\text{critical}}$. 
%However, as our solution is constructed by hand and not by following any instability. 
%What is more, we cannot see the bulk Fermi
The transition is first order, which means that the there is no continuous exchange of charge from the RN solution to the bulk Fermi sea.
%sea acquire the charge continuously, 
Instead all the charge is carried 
%even 
by the infinitesimally small rQES. 
%at birth. 
This is somewhat 
%disheartening: 
different from the conundrum that we mention in the Introduction: the %different values of the
onset of a log-oscillatory signal in the spectral function signaling a putative instability and the presence of normalizable solutions. 
%we have moved in an unexpected direction, hoping to describe how
%the \rn horizon loses charge to a Fermi-liquid Fermi surface. 
%Instead, we ended up with a rQES which 
The first order transition is essentially unrelated to the RN horizon instability.

%Therefore, 
Although it is not yet clear how the rQES 
%misses the most exciting 
%does not 
is related to the final state after the conjectured continuous quantum phase transition which destroys the \rn black hole horizon signalled by the log-oscillatory instability,
%correspond to the second order
%quantum critical point and the process which destroys the \rn black hole horizon, 
%Nevertheless, 
we nevertheless feel it is a step in the right direction, bringing us closer to the full unregulated quantum electron star.
The reasons are the following:

\begin{enumerate}
 \item It is now much clearer what a healthy Fermi liquid should do on the gravity side: it should self-consistently form a geometry which yields such an effective potential for the Dirac fermion that it is just 
 %localized 
 confined
 enough not to diverge in far IR
 but not so much that the bulk Fermi sea dies out in the far IR, failing to influence the low-energy physics.
 
 \item We have inspected in some detail the spectrum and the phenomenology of the dual confined Fermi liquid. Although our confining bulk construction is somewhat more natural in holography -- it just uses a non-dynamical rather than a dynamical scalar --- than the hard-wall model \cite{Sachdev}, and it now allows us to compute the backreaction, qualitatively the
 field-theory side description is only marginally improved. Similar to the hard-wall model, the occupied fermions have vanishing self-energy. The main effect of the backreaction is to understand how this confined Fermi gas emerges in an RG-flow from the UV conformal field theory.
 In the likely event that an unregulated (confining) quantum electron star --- supported for instance by a dynamical rather than a non-dynamical scalar (such as the fluid electron star in \cite{HartnollHuijse}) --- has a Lifshitz IR rather than an \ads IR, possible decay into the Lifshitz horizon could provide a finite lifetime and an honest Fermi liquid. 
 %This is
 %actually just the same as for the non-Fermi liquids of \cite{Faulkner:2009}.
 
 \item Unlike the global AdS radius regulator of \cite{McGreevy:2013} which cannot be easily sent to infinity, our scalar can at least in principle be made dynamical. That would be a perfectly natural holographic model, given the ubiquity of non-minimally 
 coupled scalars in top-down holographic actions. Therefore, a very natural line of further research is to turn this construction into a fully dynamical Einstein-Maxwell-Dirac-scalar system, similar to the fluid approach of \cite{HartnollHuijse}.
 %This is the most natural line of further work.
\end{enumerate}

%Thanks to the calculation of the spectrum and also some analytical insight we have in our model, we are able to draw some general conclusions about the field theory dual of the regulated QES. Classical electron star is famous for its
%infinite tower of exponentially narrow Fermi surfaces \cite{Hartnoll:es,Leiden:2011}, usually interpreted as Fermi liquids of mesinos, i.e., gauge singlet bound states \cite{HartnollHuijse,Huijse:2011ef}, although in more general supergravity
%backgrounds they may in fact be gauginos, i.e., fundamental fermions \cite{Gubser}. In setups where the IR is cut off such as the hard-wall Fermi gas of \cite{Sachdev}, it seems that indeed bound states/color singlets are the only
%possibility: the wall gives rise to a potential well where bound states form. The scalar regulator we use is in this respect only a minimal generalization of the hard wall and the outcome is thus the same -- infinitely narrow excitations with zero self-energy, corresponding to a gas of stable quasiparticles. It is somewhat disappointing that our "holographic Fermi liquid" is just a gas. However, this is unavoidable
%as long as no horizon is present. A consistent solution for a dynamical dilaton may produce a (zero-temperature) horizon, resulting in true Fermi liquid quasiparticles instead of infinitely stable particles.

Apart from the 
natural next step
%main task 
-- making the dilaton dynamic 
%and hopefully weakening the requirement that all charge sits in the bulk Fermi sea 
-- a number of other directions of work open up. It would be useful to understand the relation of our work to
the AdS/QCD studies, some of which employ a similar type of scalar (soft wall) to impose confinement. The role of the Fock correction (the one-loop exchange diagram) is also not clear yet, and may be important for a fully self-regulating solution and/or a finite self-energy. Finally, the 
most characteristic property of rQES -- the domain-wall-type solution with an infrared AdS$_4$, is analogous to the domain-wall holographic superconductor solutions of Horowitz-Roberts-Gubser-Rocha \cite{GubserNellore,HorowitzRob,Horowitz}. Based on those results and the macroscopic electron star with dynamical dilaton studied in \cite{HartnollHuijse}, it strongly suggests that Lifshitz IR quantum electron stars must also exist.
%We have discussed the connection briefly in the main text but it 
%would be instructive to understand in full detail why the solution fails to be stable in the fermionic case and how this reason could be mitigated.

\section*{Acknowledgements}

We thank Jan Zaanen for discussions. This reserach has made use of the excellent Sci-Hub service. This research was supported in part by the Netherlands Organization for Scientific Research (NWO), the Netherlands Organization for Scientific
Research/Ministry of Science and Education (NWO/OCW), Ministry of Education, Science and Technological Development of the Republic of Serbia and Science Fund of the Republic of Serbia, under the Key2SM project (PROMIS program, Grant No. 6066160).

%\clearpage

\bibliographystyle{unsrt}
\bibliography{references}

\begin{thebibliography}{10}

\bibitem{Huijse:2011hp}
Liza Huijse and Subir Sachdev.
\newblock {Fermi surfaces and gauge-gravity duality}.
\newblock {\em Phys. Rev. D}, 84:026001, 2011.

\bibitem{Vegh:2009}
Hong Liu, John McGreevy, and David Vegh.
\newblock {Non-Fermi liquids from holography}.
\newblock {\em Phys. Rev. D}, 83:065029, 2011.

\bibitem{Leiden:2009}
Mihailo \v{C}ubrovi\'c, Jan Zaanen, and Koenraad Schalm.
\newblock {String Theory, Quantum Phase Transitions and the Emergent
  Fermi-Liquid}.
\newblock {\em Science}, 325:439--444, 2009.

\bibitem{Faulkner:2009}
Thomas Faulkner, Hong Liu, John McGreevy, and David Vegh.
\newblock {Emergent quantum criticality, Fermi surfaces, and AdS(2)}.
\newblock {\em Phys. Rev. D}, 83:125002, 2011.

\bibitem{Hartnoll:es}
Sean~A. Hartnoll and Alireza Tavanfar.
\newblock {Electron stars for holographic metallic criticality}.
\newblock {\em Phys. Rev. D}, 83:046003, 2011.

\bibitem{Larus}
V.~Giangreco~M. Puletti, S.~Nowling, L.~Thorlacius, and T.~Zingg.
\newblock {Holographic metals at finite temperature}.
\newblock {\em JHEP}, 01:117, 2011.

\bibitem{Hartnoll:phtr}
Sean~A. Hartnoll and Pavel Petrov.
\newblock {Electron star birth: A continuous phase transition at nonzero
  density}.
\newblock {\em Phys. Rev. Lett.}, 106:121601, 2011.

\bibitem{Hartnoll:2011}
Sean~A. Hartnoll, Diego~M. Hofman, and David Vegh.
\newblock {Stellar spectroscopy: Fermions and holographic Lifshitz
  criticality}.
\newblock {\em JHEP}, 08:096, 2011.

\bibitem{Leiden:2011}
Mihailo \v{C}ubrovi\'c, Yan Liu, Koenraad Schalm, Ya-Wen Sun, and Jan Zaanen.
\newblock {Spectral probes of the holographic Fermi groundstate: dialing
  between the electron star and AdS Dirac hair}.
\newblock {\em Phys. Rev. D}, 84:086002, 2011.

\bibitem{thebook}
Jan Zaanen, Ya-Wen Sun, Yan Liu, and Koenraad Schalm.
\newblock {\em {Holographic Duality in Condensed Matter Physics}}.
\newblock Cambridge Univ. Press, 2015.

\bibitem{Hartnoll:2016apf}
Sean~A. Hartnoll, Andrew Lucas, and Subir Sachdev.
\newblock {Holographic quantum matter}.
\newblock 12 2016.

\bibitem{Leiden:2013}
Mariya~V. Medvedyeva, Elena Gubankova, Mihailo \v{C}ubrovi\'c, Koenraad Schalm,
  and Jan Zaanen.
\newblock {Quantum corrected phase diagram of holographic fermions}.
\newblock {\em JHEP}, 12:025, 2013.

\bibitem{Sachdev}
Subir Sachdev.
\newblock {A model of a Fermi liquid using gauge-gravity duality}.
\newblock {\em Phys. Rev. D}, 84:066009, 2011.

\bibitem{McGreevy:2012}
Andrea Allais, John McGreevy, and S.~Josephine Suh.
\newblock {A quantum electron star}.
\newblock {\em Phys. Rev. Lett.}, 108:231602, 2012.

\bibitem{McGreevy:2013}
Andrea Allais and John McGreevy.
\newblock {How to construct a gravitating quantum electron star}.
\newblock {\em Phys. Rev. D}, 88(6):066006, 2013.

\bibitem{HorowitzRob}
Gary~T. Horowitz and Matthew~M. Roberts.
\newblock {Zero Temperature Limit of Holographic Superconductors}.
\newblock {\em JHEP}, 11:015, 2009.

\bibitem{Gubser:2008wz}
Steven~S. Gubser and Fabio~D. Rocha.
\newblock {The gravity dual to a quantum critical point with spontaneous
  symmetry breaking}.
\newblock {\em Phys. Rev. Lett.}, 102:061601, 2009.

\bibitem{GubserNellore}
Steven~S. Gubser and Abhinav Nellore.
\newblock {Ground states of holographic superconductors}.
\newblock {\em Phys. Rev. D}, 80:105007, 2009.

\bibitem{HartnollHuijse}
Sean~A. Hartnoll and Liza Huijse.
\newblock {Fractionalization of holographic Fermi surfaces}.
\newblock {\em Class. Quant. Grav.}, 29:194001, 2012.

\bibitem{Herzog:2006ra}
Christopher~P. Herzog.
\newblock {A Holographic Prediction of the Deconfinement Temperature}.
\newblock {\em Phys. Rev. Lett.}, 98:091601, 2007.

\bibitem{deTeramond:2011qp}
Guy~F. de~Teramond and Stanley~J. Brodsky.
\newblock {Excited Baryons in Holographic QCD}.
\newblock {\em AIP Conf. Proc.}, 1432(1):168--175, 2012.

\bibitem{Karch:2006pv}
Andreas Karch, Emanuel Katz, Dam~T. Son, and Mikhail~A. Stephanov.
\newblock {Linear confinement and AdS/QCD}.
\newblock {\em Phys. Rev. D}, 74:015005, 2006.

\bibitem{Fang:2016uer}
Zhen Fang, Danning Li, and Yue-Liang Wu.
\newblock {IR-improved Soft-wall AdS/QCD Model for Baryons}.
\newblock {\em Phys. Lett. B}, 754:343--348, 2016.

\bibitem{Iizuka:2011hg}
Norihiro Iizuka, Nilay Kundu, Prithvi Narayan, and Sandip~P. Trivedi.
\newblock {Holographic Fermi and Non-Fermi Liquids with Transitions in Dilaton
  Gravity}.
\newblock {\em JHEP}, 01:094, 2012.

\bibitem{Landau9}
Lev~Davidovich Landau and Evgeny~Mikhailovich Lifshitz.
\newblock {\em {Statistical Physics, Part 2: Theory of the Condensed State}}.
\newblock Fizmatlit, 2004.

\bibitem{Gubser:2009dt}
Steven~S. Gubser, Fabio~D. Rocha, and P.~Talavera.
\newblock {Normalizable fermion modes in a holographic superconductor}.
\newblock {\em JHEP}, 10:087, 2010.

\bibitem{Denef:2009yy}
Frederik Denef, Sean~A. Hartnoll, and Subir Sachdev.
\newblock {Quantum oscillations and black hole ringing}.
\newblock {\em Phys. Rev. D}, 80:126016, 2009.

\bibitem{Hashimoto:2012ti}
Koji Hashimoto and Norihiro Iizuka.
\newblock {A Comment on Holographic Luttinger Theorem}.
\newblock {\em JHEP}, 07:064, 2012.

\bibitem{Liu:2011}
Nabil Iqbal, Hong Liu, and Mark Mezei.
\newblock {Semi-local quantum liquids}.
\newblock {\em JHEP}, 04:086, 2012.

\bibitem{Iqbal:2009fd}
Nabil Iqbal and Hong Liu.
\newblock {Real-time response in AdS/CFT with application to spinors}.
\newblock {\em Fortsch. Phys.}, 57:367--384, 2009.

\bibitem{Horowitz}
Gary Horowitz, Albion Lawrence, and Eva Silverstein.
\newblock {Insightful D-branes}.
\newblock {\em JHEP}, 07:057, 2009.

\end{thebibliography}

\appendix

%\section{AdS/QCD model}
%\textbf{Message of this section: add a dilaton to make the hard wall soft. Reproduces the Sachdev model but once again the backreaction on the metric does not work out. Hence the necessity to go back to a more regulated model.}

%In this section, we will introduce an EMD background which reproduces the results of the hard-wall Fermi liquid model \cite{Sachdev} without the need of any type of regulator. This model will serve as a pedagogical example into how a
%scalar field can localize the fermions in the bulk and define a confined Fermi liquid. However, as we shall explain, the backreaction on the metric is not well-controlled in this setup (just like in the original model \cite{Sachdev}
%where the presence of the hard wall makes it impossible to impose meaningful boundary conditions if one tries to solve Einstein equations). This unregulated model thus comes short of being a full-fledged quantum electron star itself,
%but motivates and paves the way to the confined quantum electron star introduced in the main text.

%\subsection{Model}
%\input{sections/IHQCD}

%\section{First-order analytical solution}

%\input{sections/swqes_analytic}

\end{document}